\documentclass[10pt]{article}

\usepackage{amsfonts}
\usepackage{amsmath}
\usepackage{amssymb}
\usepackage{compozition}
\usepackage{bm}
\usepackage{graphicx}
\usepackage[all]{xy}
\usepackage{MnSymbol}
\usepackage{comment}

\usepackage{url}
\usepackage{verbatim}
\makeatletter
\g@addto@macro{\UrlBreaks}{\UrlOrds}
\makeatother

\usepackage{tocloft}
\cftsetindents{subsection}{0em}{2.5em}  

\title{\textbf{Counting Risk Increments to Make \\ Decisions During an Epidemic}}

\author{Lucien Hardy\footnote{lhardy@perimeterinstitute.ca}\\
\textit{Perimeter Institute,}\\
\textit{31 Caroline Street North,}\\
\textit{Waterloo, Ontario N2L 2Y5, Canada}}
\date{}

\begin{document}

\maketitle

\begin{abstract}
I propose a smartphone app that will allow people to participate in the management of their own safety during an epidemic or pandemic such as COVID-19 by enabling them to view, in advance, the risks they would take if they visit some given \emph{venue} (a cafe, the gym, the workplace, the park, \dots) and, furthermore, track the accumulation of such risks during the course of any given day or week.

This idea can be presented to users of the app as counting points.  One point represents some constant probability, $p_\text{point}$, of infection.   Then the app would work in a similar way to a calorie counting app (instead of counting calories we count probability increments of being infected).  Government could set a maximum recommended number of daily (or weekly) points available to each user in accord with its objectives (bringing the disease under control, allowing essential workers to work, protecting vulnerable individuals, \dots).    It is posited that this, along with other proposed \lq\lq levers" would allow government to manage a gradual transition to normalcy.

I discuss a circuit framework with wires running between  boxes.   In this framework the wires represent possible sources of infection, namely individuals and the venues themselves (through deposits of pathogens left at the venue).  The boxes represent interactions of these sources (when individuals visit a venue).   This circuit framework allows (i) calculation of points cost for visiting venues and (ii) probabilistic contact tracing.

The points systems proposed here could complement existing contact tracing apps by adding functionality to permit users to participate in decision making up front.
\end{abstract}

\newpage

\tableofcontents

\part*{Introduction}
\addcontentsline{toc}{part}{Introduction}


\noindent
To make decisions in dangerous environments we need some way of assessing the risk corresponding to different choices we might make.  In the context of an epidemic, or a pandemic, such as COVID-19 there are two points of view that can be considered. The first point of view is that of government which wants to minimise the spread of the disease globally in accord with various objectives, such as ensuring:  that fewer vulnerable people get the disease, that the health system is not overwhelmed, essential services can continue to function and, in the long run, that the economy recovers.   The second point of view is that of individuals, or small collections of individuals (such as families or groups of friends) who want to minimise their risk of getting COVID-19, and (for the altruistic) minimise their contribution to the global spread of the disease.   Individuals also want to be able to use essential services, and look after their economic well-being.  Given sufficient assistance,  individuals and small groups of individuals are capable of quite complex decision making behaviours to navigate dangerous situations.  For example, a large percentage of the population can drive, and even those that cannot are, by and large, capable of behaving safely around vehicles.  Physical (social) distancing measures that are typically considered are rather coarse-grained  \cite{wilder2020isolation}.   Examples include closing schools, asking people to go for only one walk per day, implementing a work from home policy, etc.  Such policies are appropriate when we are in a near lock-down state.  It would be good if government had its hands on \lq\lq levers" enabling them to transition smoothly from lock-down to full-liberty state where people can move around as they wish (once the disease has ceased to be a concern).  The proposal in this paper would allow this smooth transition harnessing the  ability and intent of individuals to navigate the dangers of a pandemic by putting in place a framework individuals could use to make informed decisions concerning the risks associated with different actions.  Further, this framework could be integrated with government objectives.

The concrete proposal to achieve this is a participatory physical distancing smartphone app described in Part \ref{part:paticipatory} of this paper.  This app would allow people to view, in advance, the risks associated with different choices (such as whether to go to a particular cafe or not).  It does this by calculating a cost in terms of points.  Each point corresponds to a probability increment, $p_\text{point}$, for getting infected with the disease.   Government can set a certain recommended number of points per day, per week or per fortnight for individuals (this may depend on various factors such as age and underlying health conditions).  The points costs associated with the actual activities a person engages in are tallied.  This is rather like counting calories.  There are numerous venues where people can interact such as a dwelling, a workplace, the gym, or a supermarket.   Venues could be managed or unmanaged.  For managed venues there is a nominated person who would be tasked with entering information into an installation of  the app, perhaps with the objective of keeping the rate at which transmission of the disease occurs at that venue is below some recommended maximum.

In Part \ref{part:circuitmodel} of this paper I will put in place a mathematical framework for doing quantitative calculations based on the information collected by the app.  This will be a circuit framework consisting of boxes  with systems passing  between them (the mathematics is borrowed from work done in Quantum Foundations \cite{hardy2013formalism, hardy2016operational, chiribella2010probabilistic,  coecke2017picturing}).  Here is an example of a small fragment of a circuit describing the behaviour of three individuals visiting two venues
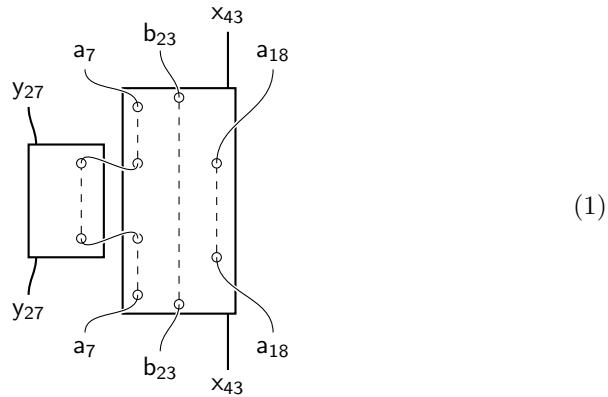
\begin{equation}  
\begin{Compose}{0}{0}  \setsecondfont{\mathsf} 
\crectangle{A}{3}{6}{0,0}   \crectangle{B}{2}{3}{-6,0}
\thispoint {xi}{2.6,-9} \csymbolalt[0,-25]{x_{43}} \jointbnoarrowthick{xi}{0}{A}{2.6}
\thispoint {xo}{2.6,9} \csymbolalt[0,25]{x_{43}} \joinbtnoarrowthick{xo}{0}{A}{2.6}
\thispoint {xiB}{-8,-5} \csymbolalt[0,-25]{y_{27}} \jointbnoarrowthick{xiB}{0}{B}{-1.6}
\thispoint {xoB}{-8,5} \csymbolalt[0,25]{y_{27}} \joinbtnoarrowthick{xoB}{0}{B}{-1.6}
\thispoint{a7start}{-5,-7}\csymbolalt[0,-25]{a_7}  \thispoint{a7in}{-2.2,-5} \jointbnover{a7start}{0}{a7in}{0}
\thispoint{a7out}{-2.2,5}\thispoint{a7end}{-5,7} \csymbolalt[0,25]{a_7}
\jointbnover{a7out}{0}{a7end}{0}
\thispoint{a7leave}{-2.2, -2}  \thispoint{a7return}{-2.2,2}
\linedashedcircles{a7in}{a7leave} \linedashedcircles{a7return}{a7out}
\thispoint{a7Bin}{-5.2,-2} \thispoint{a7Bout}{-5.2,2} \jointbnover{a7leave}{0}{a7Bin}{0}  \jointbnover{a7Bout}{0}{a7return}{0} \linedashedcircles{a7Bin}{a7Bout}
\thispoint{a18start}{5,-7}\csymbolalt[0,-25]{a_{18}}  \thispoint{a18in}{2,-3} \jointbnover{a18start}{0}{a18in}{0}
\thispoint{a18out}{2,2}\thispoint{a18end}{5,7}  \csymbolalt[0,25]{a_{18}}
\jointbnover{a18out}{0}{a18end}{0}
\linedashedcircles{a18in}{a18out}  
\thispoint{a23start}{-1,-8}\csymbolalt[0,-25]{b_{23}}  \thispoint{a23in}{0,-5.5} \jointbnover{a23start}{0}{a23in}{0}
\thispoint{a23out}{0,5.5}\thispoint{a23end}{-1,8} \csymbolalt[0,25]{b_{23}}
\jointbnover{a23out}{0}{a23end}{0}
\linedashedcircles{a23in}{a23out}  
\end{Compose}
\end{equation}
Time runs up the page.  This shows how individual $7$ visits arrives at venue $43$, leaves to visit venue $27$ then returns to venue $43$ before finally leaving.   Meanwhile, individuals $23$ and $18$ arrive at venue $43$ at different times and leave at different times.  This diagram is descriptive but it can be used to obtain a mathematical object, namely a tensor which captures the pertinent probabilities of the individuals and the venues transitioning between internal states (hidden variables) as they interact.  The key aspect of this map from the descriptive to the mathematical is that it preserves the compositional form - calculations can be represented in diagrammatic form by pictures having the same composition as the descriptive pictures.
Ideally, every system is maintained by data entry into an installation of the app though we can also accommodate unmanaged systems in the framework.  The systems  correspond to individual people (e.g.\  individuals $\mathsf{a_7}$ and $\mathsf{a_{18}}$ using the app and individual $\mathsf{b_{23}}$ not using it) and to venues  (e.g.\ $\mathsf{x_{43}}$ managed by the app and unmanaged venue $\mathsf{y_{27}}$).   Examples of a venue are a family home, a cafe, a sidewalk, a work place, a supermarket\dots.   The boxes correspond to social interactions (meeting a friend for coffee in a cafe, going to work, a visit to the supermarket\dots).  I will consider strategies for using this mathematical framework to inform the points costs for different social activities.   Symptoms and positive or negative test results for the disease can be entered using the app interface.   Using this mathematical framework, the app can provide the user with an ongoing report of the probability they have contracted the disease.   This probability typically will be low if the user has stuck to the recommended points allocation, though if somebody they have come into contact with subsequently reports symptoms or a positive test result, their probability will increase.  In this way, the framework can accomplish what might be termed \emph{probabilistic contact tracing}.  People with probability of infection above some threshold can be asked to self-isolate.   It will cost more points to interact with people who have spent more points (who typically have a higher probability of having contracted the disease even if they fall below the threshold at which people are instructed to self isolate) and it is posited that this will help suppress spread of  the disease.

The app being proposed involves the notion of agency (users make choices).  It is difficult to incorporate agency into physical models.  My own research (e.g.\ \cite{hardy2013formalism, hardy2016operational}) and that of many of my colleagues (e.g.\ \cite{chiribella2010probabilistic, leifer2013towards, coecke2017picturing, ried2019modelling, oreshkov2012quantum}) concerns agency in fundamental physics (such as Quantum Theory and General Relativity).   Agent based approaches have also been taken in other more relevant fields. For example, in the context of understanding the large scale impact of human actions in the context of behavioural science (see, for example, \cite{grimm2005individual}).

Contact tracing \cite{huerta2002contact} is a tool to find and isolate infected individuals reducing the extent to which they infect others.
Numerous contact tracing apps have been developed, or are under development particularly in the wake of the COVID-19 crisis.  Here is a very incomplete list \cite{2003.08567, tracetogether2020, aarogya2020, covidshield2020, nhscovid19app2020, alsdurf2020covi}.  See \cite{phonestrack2020, googlemay2020} for some popular articles.  Google and Apple have provided notification technology to assist contact tracing app developers \cite{googleapple2020} while enabling privacy through cryptographic techniques. There has been much discussion of ethics (for example see \cite{gasser2020digital} and the statement \cite{jointcivilsociety2020} on Amnesty International's website co-signed by numerous relevant bodies).  

Contact tracing apps tell people after the fact that they have come into contact with somebody who has (or might have had) the disease.   There has been very little discussion of frameworks to allow individuals to make decisions up front as proposed here.  The COVI project \cite{alsdurf2020covi} led by Bengio is developing an app that uses machine learning and epidemiological modeling.  They discuss empowering individuals by providing a \lq\lq tailored recommendations feature that helps users make real-time decisions daily about their activities based on their personal level of risk".  They also discuss probabilistic contact tracing.  Wu discusses networks for small groups of individuals that optimise against the spread of the disease in a game theory context \cite{wu2020social}.  Popa \cite{popa2020decision} discusses automated tools that make recommendations to local government on physical distancing measures (such as which schools and factories to close).
 
Epidemiological models for the spread of disease in a large community (a country, state, province, or city for example) require very large computational resources that are well beyond what is possible on a smartphone, computer, or even readily available in the cloud for large numbers of people to use in real time (for example, see \cite{ferguson2020impact}).  There is, then, a need to design the infrastructure for the proposed system such that the computations required are feasible in real time.   Machine learning \cite{Nielsen2015, Goodfellow-et-al-2016, alpaydin2020introduction} is a rapidly developing field with a wide range of applications across industry and may have applications to the proposal considered here.

\part{A participatory physical distancing app}\label{part:paticipatory}

For the sake of definiteness, it is worth considering one version of how a participatory physical distancing app would work.  Many variations on this idea are possible.   In the proposal below each person  participating will manage an \emph{individual account} through the app.  Venues are either managed or unmanaged.  For managed venues, a nominated person for every venue using the app will manage a \emph{venue account} using the app and maintain certain standards.   Unmanaged venues can be given a label so they can be identified by individuals visiting them but are not managed.    Examples of venues that might be managed are a work place, a train carriage, a cafe, a supermarket, a taxi, and a small park.  Examples of venues that might be unmanaged are a large park, footpaths and sidewalks.

\section{Individual accounts}

Individual accounts have the following features.
\begin{description}
  \item[Questionnaire.] On opening an individual account, the user enters pertinent information like age,  underlying health conditions, and occupation (such as whether the individual is working in essential services).
  \item[Recommended points allocation.] A recommended daily points allowance is calculated for the user.
  \item[Daily tracking.] Activities like going to the supermarket, meeting a friend, etc, are tracked.   The tracking can be done manually and be assisted by the GPS function of the smartphone, or it could be largely automated.
  \item[Costs.] Each activity the user may engage in costs a certain number of points.  An activity costing $N$ points corresponds roughly to a probability of $Np_\text{point}$ of contracting the disease when engaging in that activity  (where $p_\text{point}$ is some constant probability).  The number of points can be viewed beforehand to help the user decide if they want to engage in the activity.
  \item[Rating.]  A user will be awarded a higher rating for tracking consistently.  The calculation of the rating could be coupled with GPS data collected by the smartphone to check the user is tracking.
  \item[Health reporting.]  The app asks the user to report any changes in health that have a bearing on the disease (such as developing symptoms, or a positive or negative test result for the disease).
  \item[Reporting.]  The app will report to the user what their probability of having contracted the disease is (according to the mathematical modeling).  The app will issue an alert if this rises above some value at which the individual should self isolate.
\end{description}
Some people (children, some elderly people, people without smartphones, and so on) are incapable of keeping an account by themselves but someone else may maintain an account on their behalf.   There will also be people who do not have an account at all - such people need to be taken into account in the modelling.

\section{Behaviours, trajectories, and paths}

Here we distinguish three levels of description of what people might do when in a venue.

A \emph{behaviour} is something a person can reasonably choose to do and, therefore, has control over.  Alice can choose to arrive at a cafe between 3:30PM and 3:35PM with her friend Bob, they can queue to order coffee and snacks, then sit at table 8 consume the coffee and snacks after they are delivered by the server, and leave between 4:30PM and 4:35PM.  This description is coarse-grained.

A more fine-grained description of what Alice does might be available from the Bluetooth features on her phone.  We will call this more fine-grained level of description a \emph{trajectory}.   The information provided by Bluetooth may be relative rather than absolute (for example, it may specify how close Alice is to the server to some level of accuracy without specifying their absolute positions).

We can also consider the actual \emph{path} taken by Alice where this is specified exactly.  There is no instrumentation in place to measure this but it is a useful concept in calculations.  In particular, behaviours and trajectories can each be thought of as probability distributions over paths.

Alice may decide on a certain behaviour.  This is consistent with many trajectories.  If the Bluetooth function is turned on then there is more fine-grained information available going beyond what we might reasonably expect Alice to choose.  This means that we can calculate \lq\lq fine-grained" probabilities (depending on the trajectory information).   This raises an important issue.  Should we use the fine-grained probability or the \lq\lq coarse-grained" probability (determined from behaviour) when we calculate the number of points it costs to visit a venue?  Since we are endeavouring to provide people with a behavioural framework which influences the way they make choices, it is better to base the number of points on the coarse-grained probabilities based on behaviours (which people have control over).  Nevertheless, the app can be designed to provide a more accurate probabilistic picture should users be interested.   If the costing is done appropriately, then the average course-grained probability should equal the average fine-grained probability when averaged over many visits to venues.

\section{Managed venues}

Venue accounts are run by a nominated manager.  Their task is to track visitors to the venue so it is possible to quantify the risk of infection for any given visitor.  We use the term \lq\lq visitor" to include any person who enters the venue. For a store, for example, this includes customers, and people working at the store.  Risks come from other people physically present at the venue at the same time and from the virus left at the venue (on surfaces or airborne) by visitors who may have since left.  Managed venue accounts have the following features.
\begin{description}
  \item[Questionnaire.] On opening a venue account, the nominated venue manager will answer various questions about the venue that can be used to calculate the extent to which that venue can lead to transmission of the disease.  This may include details such as square footage, air circulation,  typical distances between people, the areas of surfaces that people may touch.      There may also be questions about the purpose of the venue (does it constitute essential services for example).
  \item[Recommended procedures.]   The app can recommend procedures which will reduce the chance of transmission of the disease.  For example, a supermarket could limit the number of people allowed in at any given time and wipe down surfaces every hour.   The app could issue a notification to follow a cleaning procedure when it detected that the probability for disease transmission through contact with exposed surfaces in the venue was bigger than a certain threshold.
  \item[Behaviours.]   Each venue provides a set of behaviours available to visitors to choose from.
  \item[Recommended cost rate reporting.]   The cost rate (points cost per hour) is equal to the infection rate (for a typical individual) divided by $p_\text{point}$.   The cost rate for a given behaviour at a venue at any given time will depend on who is visiting it, or has visited it recently enough to leave pathogen deposits behind.  This rate can be estimated by the app.   Venues can adopt one of two strategies.  Either they can simply report the current cost rate so that visitors can decide whether or not they wish to visit the venue at that given time.  Or a venue can strive to keep the cost rate below a certain recommended threshold.
    \item[Tracking.]  The venue will track people coming through according in order to  estimate the infection rate.   This tracking information can also be used for probabilistic contact tracing.   The venue may, according to its remit, refuse or delay entry to limit spread of the disease.
  \item[Rating.]  Places that track consistently and indicate they follow recommended procedures will get a higher rating. GPS and Bluetooth on smartphones could be used to check that people visiting the venue are being tracked.  Feedback from users visiting the venue could be used to indicate that procedures are being carried out.
  \item[Reporting.]  The app will provide reporting to the venue manager on parameters relating to the risks visitors face.
\end{description}

\section{Unmanaged venues}

It is impractical for all places to be managed.   We can envisage various approaches to such unmanaged venues.  Even if no individual is nominated to manage a particular venue, that venue can still have a name (drawn from map data) and we can still envisage people tracking when they are visiting the venue.  Under such circumstances, potential visitors at a given time could obtain information about current and recent usage to enable them to decide whether to visit.   Costs for such visits could be calculated from generic cost functions for such types of venue.   This kind of approach would be well suited to spacious outdoor spaces.   For example, the cost for a picnic with a friend in a park would depend on the recent activity of that friend and be calculated by some function taking into account that this is an outdoor space.   In other cases, like a poorly ventilated building, unmanaged venues will represent an unquantifiable risk and people may decide not to visit.

\section{How it works}\label{sec:howitworks}

To illustrate how this all works consider an example of two friends, Alice and Bob, meeting at a cafe for lunch.   Each friend will, beforehand, use the app to see how many points this will cost them.   They each enter the identifier of the cafe and also of the friend into the app.  Behaviours are allocated and the app does a calculation telling both Alice and Bob the estimated points cost for this. If the points costs are acceptable to both of them and the cafe is happy to accept the risk, then they log their intention, go to the cafe and have lunch.  Afterwards, the app does a calculation of the actual points cost - taking into account the actual behaviours (for example, they may stay a little longer than they had intended).   These actual points costs are debited against Alice and Bob's points allocation.

Most likely, the cost for Alice will depend on Bob's previous behaviour to a greater extent than the behaviour of others because she is sitting close to him for a longer period of time.  This means that Bob is motivated to stick to the recommended points allocation.  Similar remarks apply to Alice.  Thus, there is a negative feedback loop that may help control the spread of the disease.  

A simplified approach to calculating points is discussed in Appendix \ref{app:simplifiedpointscalc}.  This involves using a weighted sum over the costs for previous visits to venues of people in the cafe.   This simplified approach may be useful but it relies on various assumptions that may not be valid.  The best way to calculate the cost to Alice for visiting the cafe is to do a full calculation of the probability increment for Alice on visiting the cafe without these assumptions (and similarly for Bob).  This can be done using the circuit  techniques developed in Part \ref{part:circuitmodel} (see in particular the  Sec.\ \ref{sec:calculatingprobabilitiesandpoints}).

\section{Predicted and actual costs}\label{sec:predictedandactualcosts}

We can be more nuanced about costs and probabilities by distinguishing the following
\begin{description}
  \item[Predicted cost.]   This is the predicted cost for the visit  based on the anticipated behaviour of the user.
  \item[Actual cost.]  This is the actual cost for the visit to the venue based on the actual behaviour.  For example, the user may stay longer than they had anticipated so this would increase the cost.
  \item[Fine-grained probability.]  There may be additional information providing a more fine-grained picture of the individuals behaviour as provided by GPS and bluetooth features. Furthermore, there may be more accurate information about the other visitors to the venue.  This may make it possible to do a more accurate calculation of the probability of infection during the visit.
  \item[Updated probability.]  New information may subsequently become available allowing the probability of infection to be updated.  For example, if somebody else at a venue the individual visits subsequently tests negative for the disease then their updated probability of infection may be lower than before.
\end{description}
A natural question is how should these different costs and probabilities be incorporated into the app?  One proposal is the following.  The app lets the user know the predicted costs beforehand.  Then the app charges the user the actual cost after their visit (so this is subtracted from the recommended daily (or weekly) number of points).   The app, further, allows reporting of the fine-grained probability to the user and updates this fine-grained probability when new information becomes available.    If this updated fine-grained probability rises above various threshold values then various policies may be implemented.   Possible policies associated with different thresholds are (i) reduction of allocated points allowance, (ii) increase in coefficients that determine the costs for others to interact with this individual, and (iii) the user is asked to self-isolate.   It is worth noting that subsequent information may be made available that may reduce the updated probability. This could happen if, for example, some other person, $m$, reports that they are infected causing individual $n$'s probability to go up. But then another person, $m'$, could report they are infected suggesting a different pathway to $m$ having become infected so that individual $n$'s probability will go down again.  In such a circumstance, an instruction to self-isolate could be retracted.

If calculations are done properly then the actual cost multiplied by $p_\text{points}$ should equal the average of the fine-grained probability for visits to the venue. If this is not the case, there is a problem with the model and we could adjust the parameters by which the actual cost is calculated.  This could be implemented by a machine learning algorithm.  Further information that might be incorporated into a machine learning algorithm could come from reports of symptoms and test results that relate to the given venue.   No technical discussions of the use of machine learning are given in the current paper.

\section{Levers for government}

Before looking into how to calculate probabilities, it is useful to see how government agencies could use this app to manage their response to the epidemic or pandemic.
\begin{description}
  \item[Set recommended number of points.]  The recommended number of points is a way to put an upper limit on an individuals contribution to the probability of infection at a venue (since this will depend on their recent interactions before visiting the given venue).  At different levels of lockdown government agencies may want to be more or less lenient.   Also, the government may need set higher recommended numbers of points for essential workers and a lower number for individuals that are more vulnerable to the disease.
  \item[Costs.] While the costs for different activities would ordinarily be calculated by the algorithms provided with the app, government agencies could artificially increase these costs to control the flow of people for different ends.   Additionally, there is a role for government in setting the standards by which these costs are calculated.
  \item[Threshold probability for isolation.]  Government would set the threshold probability above which people are instructed to self-isolate.
  \item[Manage risks to essential workers.]   Health care professionals take greater risks. Nevertheless, it is counterproductive to have large numbers of essential workers falling ill.  The app could be used to take a  whole system approach to monitoring the risks to essential workers according to the best models and take remedial steps where appropriate.
  \item[Coordinate sharing of facilities.]  Some facilities are under great demand and it may make sense from an epidemiological point of view to coordinate the sharing of these facilities.  This could be done by artificially increasing the costs to segments of the population at different times during the course of the day and issuing notifications to share this information.
  \item[Establish bubbles.]  It is possible, through the costs structure, to establish bubbles \cite{block2020social} - larger groups of people - who are have minimal interaction with others.    This would be preferable to maintaining physical barriers to limit peoples mobility.   The app could provide instruction as to when it is permissible to transfer from one bubble to another.
\end{description}

\part{Circuit model}\label{part:circuitmodel}

\section{Systems and hidden variables}\label{app:systemsandhiddenvariables}

In some approaches to epidemiological modeling, graphs are used to model the whole population \cite{ferguson2020impact}.   These are very large graphs requiring vast computing resources to process, or else they are smaller graphs that coarse-grain over large groups of people.  There is another place graphs  might be used.  This is in modelling a smaller number of individuals.   Then, typically, we only have access to fragments of the graph associated with a small number of individuals.  The specific techniques I will use are adapted from operational probabilistic models as used in the study of the foundations of quantum theory.   

Individuals are, at all times, in some place (home, a cafe, work, the park, \dots) which we call venues.  We label venues by $v=1, 2, \dots$.
We will consider direct and indirect infection.  Indirect infection is where one person leaves pathogens behind at a venue (for example, on a surface, in some food, or lingering in the air)  and these infect another person at a later time.  Direct infection is when one person releases pathogens which impinge on the other person more or less immediately (for example, by hand shaking or coughing).   This division is, to some extent, arbitrary and different models may divide them up differently.   However, main concept is that indirect means of infection are delayed which suggests they can be mitigated on longer time scales by management of the venue (by cleaning surfaces for example).

We will  use a circuit to model this wherein systems pass between boxes.  We have four types of system.  We denote registered individual, $n$, by $\mathsf{a}_n$ and the managed venue, $v$, by $\mathsf{x}_v$.   Unregistered individuals are represented by $\mathsf{b}_l$ where $l$ is a temporary label applied for the given visit to the given venue.  Unmanaged venues are labeled by $\mathsf{y}_v$ where the label $v$ may be temporary or might be drawn from map information.  We could add additional systems. For example, registered individuals who do not carry a smartphone (like school children) could be denoted by $\mathsf{c}_n$.   For a disease that is spread by additional agents such as food, or animals we  might add additional system types corresponding to these agents.  In general it makes sense to add systems if the hidden variables (see below) associated with those systems are different, or the way the calculation deals with the hidden variables for those systems is different from already defined systems.

There are various quantities pertaining to an individual that are relevant to the disease we are studying such as contagiousness and  susceptibility.   We suppose these are functions of some underlying hidden variables.   We will denote the hidden variables associated with $\mathsf{a}_n$ by $a_n$, with $\mathbf{b}_l$ by $b_l$.

There are also pertinent quantities pertaining to venues such as the numbers of pathogens deposited on various surfaces, or lingering in the air.   We suppose, similarly, that these are functions of hidden variables associated with venues.   We  denote the hidden variables associated with the managed venue $\mathbf{a}_v$ by $x_v$, and with the unmanaged venue $\mathsf{y}_v$ by $y_v$.

The choice of hidden variables will depend on what it is we are trying to calculate, and how precisely we are attempting to calculate it.  For the purposes of tractability in computer modeling,  it will be necessary to discretize these hidden variables.  We will assume that $a_n$, $b_l$, $x_v$ and $y_v$, are the discretized versions of these variables.  In the framework I will set up these hidden variables will correspond to the indices in tensors and we will, consequently, be summing over them.

\section{The composition principle}

In Sec.\ \ref{sec:descriptionsusingcircuits}  I will discuss how to \emph{describe} small scale epidemiological situations in which people interact at venues using circuits (and fragments of circuits).  In the kind of approach I am proposing, description comes first.  This description can be written in diagrammatic or symbolic form.   Once the description is given, we can decide what kind of mathematical questions we wish to ask.   The description we will provide of individuals visiting venues is compositional - we wire together boxes.   The point of being careful about the compositional description is that we can convert these descriptions directly into calculations \emph{having the same compositional form}.  Indeed, the calculations can be represented by diagrams that look exactly like the diagram representing the description.  In \cite{hardy2013theory} this was called the \emph{composition principle}.    This attitude is also present in category theory where we have functors which map between categories \cite{coecke2017picturing}.

Here are two types of calculation we might consider:
\begin{description}
  \item[Probabilistic.] We may be interested in probabilities for various outcomes.  We can base this calculation on a calculation for joint distributions over the underlying hidden variables.  This framework allows us to allocate probabilities of infection to individuals,  take into account information acquired from symptoms reporting, test results, and Bluetooth proximity monitoring and also  perform probabilistic contact tracing.
  \item[Simulation.]  We may be interesting in running a simulation where we assume some particular individuals are initially infected and see, for this particular run, which individuals are infected after some time.  We can treat infection events probabilistically and consequently, if we redo the simulation we will almost certainly get different results.   These simulations could be useful to analyse the impact of different costing strategies.
\end{description}
We will show how to perform the probabilistic calculations in accord with the composition principle - that is the calculation looks like the description. These calculations can be used for calculating points costs and also implementing probabilistic contact tracing.  It is also clear that simulation calculations can also be performed in accord with the composition principle.  We will not detail how to do simulation calculations as this falls outside the remit of the present work.

The composition principle leads us to take seriously the description aspect of setting up a problem since, once we have that, the step to the calculational aspect becomes much simpler.

\section{Description using circuits}\label{sec:descriptionsusingcircuits}

In the first place I will discuss how to \emph{describe} small scale epidemiological situations using circuits (and fragments of circuits).  In the kind of approach I am proposing, description comes first.  This description can be in diagrammatic or symbolic form.   Once the description is given, we can decide what kind of mathematical questions we wish to ask.  In the approach I will present, the calculation corresponding to these mathematical questions has the same compositional form as the description (so the diagram representing the calculation looks the same as the diagram representing the description).   In Sec.\ \ref{sec:probabilisticcalculations} I will outline how to do probabilistic calculations.

\subsection{Operations}

An operation corresponds to people visiting a venue according to various behaviours.  In specifying an operation there are various \emph{settings} corresponding to choices made by the manager of the venue and the visitors to this venue. Examples of settings include (i) implementing procedures such as cleaning, (ii) the choice of behaviour of each individual visiting the venue.    Further, there can be \emph{outcomes}.  Outcomes are things that happen.   Examples of outcomes are: (i) when one of the individuals visiting the venue reports certain symptoms (through the app), (ii) when an individual receives a positive (or negative) test result for the disease, (iii) when Bluetooth information provides more detailed information about the trajectory of some or all of the individuals visiting the venue than is specified in the assigned behaviours of those individuals.   An operation is specified by providing the settings and some given outcomes.

In the graphical notation an operation is represented by a box.   Each person visiting the venue will engage in a certain behaviour.  We can partition their behaviour into temporal and spacial aspects.  The vertical dimension of the box corresponds to time and the horizontal dimension labels (though does not depict) the spacial aspects of their behaviour.   We represent an operation diagrammatically as shown in the following example
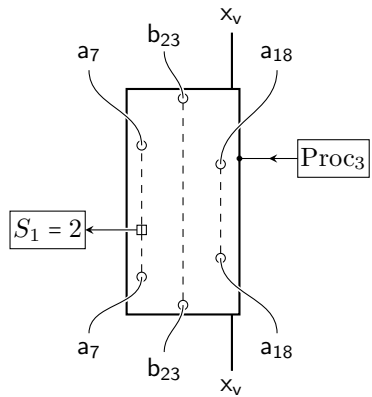
\[
\begin{Compose}{0}{0}  \setsecondfont{\mathsf} 
\crectangle{A}{3}{6}{0,0} \indashedout{A}{-2.2,-4}{-2.2,3}{7} \indashedout{A}{0,-5.5}{0,5.5}{23} \indashedout{A}{2,-3}{2,2}{18}
\relpoint{A}{2.6,-9}{vstart}\csymbolalt[0,-25]{x_v} \relpoint{A}{2.6,9}{vend} \csymbolalt[0,25]{x_v}
\relpoint{A}{-5,-7}{7start} \csymbolalt[0,-25]{a_7} \relpoint{A}{-5,7}{7end} \csymbolalt[0,25]{a_7}
\relpoint{A}{-1,-8}{23start} \csymbolalt[0,-25]{b_{23}} \relpoint{A}{-1,8}{23end} \csymbolalt[0,25]{b_{23}}
\relpoint{A}{5,-7}{18start} \csymbolalt[0,-25]{a_{18}} \relpoint{A}{5,7}{18end} \csymbolalt[0,25]{a_{18}}
\jointbnoarrowthick{vstart}{0}{A}{2.6} \jointbnoarrowthick{A}{2.6}{vend}{0}
\jointbnover{7start}{0}{A in7}{0} \jointbnover{A out7}{0}{7end}{0}
\jointbnover{23start}{0}{A in23}{0} \jointbnover{A out23}{0}{23end}{0}
\jointbnover{18start}{0}{A in18}{0} \jointbnover{A out18}{0}{18end}{0}
\rightflagdot[1.9]{A}{3,2.3}{\text{Proc}_3}{5,0}
\leftflagsquare{A in7}{0,2.5}{S_1=2}{-5,0}
\end{Compose}
\]
This has the following features.    There is a wire, $\mathbf{x}_v$, inputted at the bottom and outputted at the top corresponding to the venue itself (so time runs up the page).   Then there are wires inputted and outputted at various times corresponding to visitors who enter and leave at various times.  Shown are registered app users $\mathsf{a}_7$ and $\mathsf{a}_{18}$ and the unregistered individual $\mathsf{b}_{23}$.  The flags represent choices and outcomes.   The $\text{Proc}_3$ flag indicates that procedure 3 is carried out at a time indicated by the vertical position of the black dot. The direction of the arrow indicates this is a setting.  The $S_1=2$ flag indicates that the user reports the onset of symptom $S_1$ at level $2$ (perhaps indicating a moderate cough) at a time indicated by the vertical position of the small square.  The direction of the arrow indicates this is an outcome.    The dashed lines represent the different behaviours that have been chosen by individuals in the venue.   Ideally, we would label these behaviours with additional flags indicating that they are choices but the diagram is already quite crowded so we take this information to be implicit in the position of the dashed line.   There is additional data associated with venues and individuals that is not represented on this diagram.  For a venue this might include the physical dimension of the space and the furniture therein.  For an individual this includes any details that may bear on the development of the disease (e.g. age and health conditions).   We could include this information in the hidden variables associated with the venue and individuals (this is not done for the simple example of hidden variable allocation given in Sec.\ \ref{sec:simplifiedhiddenvariables}).   Alternatively, we can simply assume that this information is available when we perform calculations concerning a given venue and given individuals.

This diagrammatic information is quite sufficient for our purposes. However, it is sometimes instructive to refer to symbolic notation.  We represent this situation symbolically as follows
\begin{equation}
\mathsf{V}^{\mathsf{x}_\mathnormal{v}^9\mathsf{a}_7^{32}\mathbf{b}_{23}^{2}\mathsf{a}_{18}^{58}  }      _{\mathsf{x}_\mathnormal{v}^8\mathsf{a}_7^{31}\mathbf{b}_{23}^{2}\mathsf{a}_{18}^{57}}
[\text{settings}, \text{outcomes}]
\end{equation}
The superscripts on the $\mathsf{a}$'s indicate the visit number for that individual.  This number is increased by 1 after each visit. This superscript is not necessary in the diagrammatic notation as it is evident from the diagram that the visit number has incremented.  The superscript on $\mathsf{x}$ is the iteration number for the venue. It is increased by 1 for subsequent operations for that venue.  We will use the notation
\begin{equation}
\mathsf{A}^{\mathsf{x}_\mathnormal{v}^9\mathsf{a}_7^{32}\mathbf{b}_{23}^{2}\mathsf{a}_{18}^{58}  }      _{\mathsf{x}_\mathnormal{v}^8\mathsf{a}_7^{31}\mathbf{b}_{23}^{1}\mathsf{a}_{18}^{57}}
=
\mathsf{V}^{\mathsf{x}_\mathnormal{v}^9\mathsf{a}_7^{32}\mathbf{b}_{23}^{2}\mathsf{a}_{18}^{58}  }      _{\mathsf{x}_\mathnormal{v}^8\mathsf{a}_7^{31}\mathbf{b}_{23}^{1}\mathsf{a}_{18}^{57}}
[\text{settings}, \text{outcomes}]
\end{equation}
where the information in the square brackets is implicit in the symbol $\mathsf{A}$.  For a different case we use a different symbol ($\mathsf{B}$, $\mathsf{C}$, etc.\ ).  Correspondingly, we can use the following abbreviated diagrammatic notation
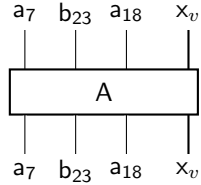
\[
\begin{Compose}{0}{0}  \setsecondfont{\mathsf} 
\crectangle{A}{5}{1.2}{0,0} \csymbol{A}
\thispoint{a1}{-4.2,-3.3}  \csymbolalt[0,-25.5]{a_7}   \jointbnover{a1}{0}{A}{-4.2}
\thispoint{d4}{-4.2,3.3}  \csymbolalt[0,25]{a_7}         \joinbtnoarrow{d4}{0}{A}{-4.2}
\thispoint{b2}{-1.5,-3.3} \csymbolalt[0,-25]{b_{23}}        \jointbnover{b2}{0}{A}{-1.5}
\thispoint{e5}{-1.5,3.3} \csymbolalt[0,25]{b_{23}}           \joinbtnoarrow{e5}{0}{A}{-1.5}
%
%
\thispoint {c3}{1.2,-3.3} \csymbolalt[0,-25]{a_{18}} \jointbnover{c3}{0}{A}{1.2}
\thispoint {f6}{1.2,3.3} \csymbolalt[0,25]{a_{18}} \joinbtnoarrow{f6}{0}{A}{1.2}
\thispoint {xi}{4.5,-3.3} \csymbolalt[0,-25]{x_\mathnormal{v}} \jointbnoarrowthick{xi}{0}{A}{4.5}
\thispoint {xo}{4.5,3.3} \csymbolalt[0,25]{x_\mathnormal{v}} \joinbtnoarrowthick{xo}{0}{A}{4.5}
\end{Compose}
\]
This obscures the times at which people arrive and leave the venue.

\subsection{Repeat visits}

A particular individual may visit a venue more than once during the time associated with a given operation.   We represent this by sending wires back after visiting one or more other  venues.  For example
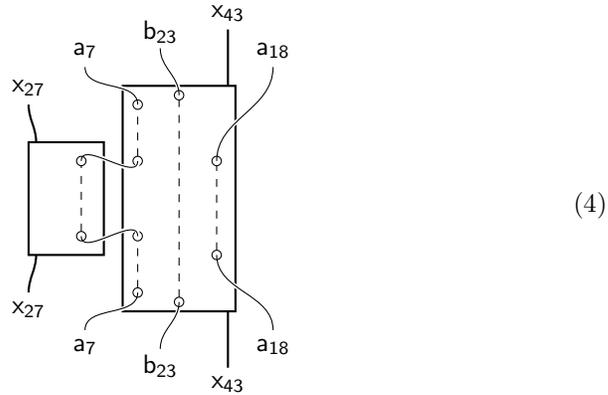
\begin{equation}\label{poppingoutdiagram}  
\begin{Compose}{0}{0}  \setsecondfont{\mathsf} 
\crectangle{A}{3}{6}{0,0}   \crectangle{B}{2}{3}{-6,0}
\thispoint {xi}{2.6,-9} \csymbolalt[0,-25]{x_{43}} \jointbnoarrowthick{xi}{0}{A}{2.6}
\thispoint {xo}{2.6,9} \csymbolalt[0,25]{x_{43}} \joinbtnoarrowthick{xo}{0}{A}{2.6}
\thispoint {xiB}{-8,-5} \csymbolalt[0,-25]{x_{27}} \jointbnoarrowthick{xiB}{0}{B}{-1.6}
\thispoint {xoB}{-8,5} \csymbolalt[0,25]{x_{27}} \joinbtnoarrowthick{xoB}{0}{B}{-1.6}
\thispoint{a7start}{-5,-7}\csymbolalt[0,-25]{a_7}  \thispoint{a7in}{-2.2,-5} \jointbnover{a7start}{0}{a7in}{0}
\thispoint{a7out}{-2.2,5}\thispoint{a7end}{-5,7} \csymbolalt[0,25]{a_7}
\jointbnover{a7out}{0}{a7end}{0}
\thispoint{a7leave}{-2.2, -2}  \thispoint{a7return}{-2.2,2}
\linedashedcircles{a7in}{a7leave} \linedashedcircles{a7return}{a7out}
\thispoint{a7Bin}{-5.2,-2} \thispoint{a7Bout}{-5.2,2} \jointbnover{a7leave}{0}{a7Bin}{0}  \jointbnover{a7Bout}{0}{a7return}{0} \linedashedcircles{a7Bin}{a7Bout}
\thispoint{a18start}{5,-7}\csymbolalt[0,-25]{a_{18}}  \thispoint{a18in}{2,-3} \jointbnover{a18start}{0}{a18in}{0}
\thispoint{a18out}{2,2}\thispoint{a18end}{5,7}  \csymbolalt[0,25]{a_{18}}
\jointbnover{a18out}{0}{a18end}{0}
\linedashedcircles{a18in}{a18out}  
\thispoint{a23start}{-1,-8}\csymbolalt[0,-25]{b_{23}}  \thispoint{a23in}{0,-5.5} \jointbnover{a23start}{0}{a23in}{0}
\thispoint{a23out}{0,5.5}\thispoint{a23end}{-1,8} \csymbolalt[0,25]{b_{23}}
\jointbnover{a23out}{0}{a23end}{0}
\linedashedcircles{a23in}{a23out}  
\end{Compose}
\end{equation}
shows how individual $7$ visits venue $43$ then venue $27$ then returns to venue $43$.   Symbolically we can write this as
\begin{equation}\label{justpoppingout}
\mathsf{A}^{\mathsf{x}_{33}^9\mathsf{a}_7^{32}\mathsf{a}_7^{34}\mathsf{a}_{18}^{58}  }      _{\mathsf{x}_{33}^8\mathsf{a}_7^{31}\mathsf{a}_7^{33}\mathsf{a}_{18}^{57}}
\mathsf{B}^{\mathsf{x}_{27}^{19}\mathsf{a}_7^{33}}_{\mathsf{x}_{27}^{18}\mathsf{a}_7^{32} }
\end{equation}
Note how the superscript $\mathsf{a}_7^{32}$ on $\mathsf{A}$ is matched with itself as a subscript on $\mathsf{B}$ corresponding to when individual $7$ leaves venue $43$ and enters venue $27$.  There is a similar repeated index $\mathsf{a}_7^{33}$ for the return trip.

\subsection{Sequential boxes for a venue}\label{sec:sequentialboxes}

Each operation corresponds to a given duration for the venue.  We can choose this duration to be longer or shorter.  We can break up longer durations into shorter durations if this is convenient.   Diagrammatically this is represented by sequential boxes for the given venue. There is, however, a subtlety.    The diagram looks like this
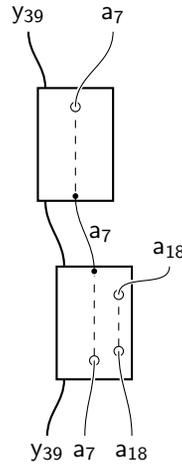
\[
\begin{Compose}{0}{0} \setsecondfont{\mathsf} 
\crectangle{A}{2}{3}{0,0} \indasheddot{A}{0,-2}{0,2.75}{7} \indashedout{A}{1.3,-1.5}{1.3,1.5}{18}
\relpoint{A}{-2.5,-6}{vstart}\csymbolalt[0,-25]{y_{39}} \relpoint{A}{-0.5,-6}{7start} \csymbolalt[0,-25]{a_7} \relpoint{A}{2,-6}{18start} \csymbolalt[0,-25]{a_{18}}
 \relpoint{A}{4,3}{18end}\csymbolalt[0,25]{a_{18}}
\crectangle{B}{2}{3}{-1,9.5} \dotdashedout{B}{0,-2.75}{0,2}{7}
\relpoint{B}{-2.5,6}{vend}\csymbolalt[0,25]{y_{39}} \relpoint{B}{2,6}{7end}\csymbolalt[0,25]{a_7}
\jointbnoarrowthick{vstart}{0}{A}{-1.6} \jointbnoarrowthick{A}{-1.6}{B}{-1.6} \jointbnoarrowthick{B}{-1.6}{vend}{0}
\jointbnover{7start}{0}{A in7}{0} \jointbnoarrow[right]{A out7}{0}{B in7}{0} \csymbolalt[10,0]{a_7}  \jointbnover{B out7}{0}{7end}{0}
\jointbnover{18start}{0}{A in18}{0} \jointbnover{A out18}{0}{18end}{0}
\end{Compose}
\]
Here individual $7$ remains in venue $39$ between the two boxes.   We use black dots to indicate that individual $7$ has remained at the venue.  The reason for doing this is that the behaviour, $u$, for individual $7$ will, in general, be specified over the whole duration of his trip to venue $39$.  In special cases, we may be able to write the behaviour as $u_1$ for the first box then $u_2$ for the second box.   Then we write $u=u_1u_2$.    However, this will not work for a general case.  Imagine venue $39$ is a store.  Individual $7$ arrives through the entrance door at a certain time.  He then wanders around the store. At the intermediate time associated with the change from the first box to the second, he may be anywhere.  Later still, he pays and leaves by the exit door.  His behaviour is described by a probability over various random walks through the store which does not factorize into separate probability distributions over the two time intervals.  In this example we cannot write the behaviour in the factorized form $u=u_1u_2$.

\subsection{Initialisation operations}

It is useful to define initialisation operations.   First, we let
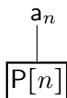
\[
\begin{Compose}{0}{0}  \setsecondfont{\mathsf}
\crectangle{n}{1.6}{1}{0,0}  \csymbol{P[\mathnormal{n}]}\thispoint{nup}{0,3}  \jointbnover{n}{0}{nup}{0}\csymbolalt[10,28]{a_\mathnormal{n}}
\end{Compose}
\]
correspond to the initialisation of individual $n$ when they first install the app.  This box can be inputted into a circuit.

We can have similar boxes for an unregistered individual at the time $t$ associated with the given output wire
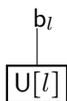
\[
\begin{Compose}{0}{0}  \setsecondfont{\mathsf}
\crectangle{n}{1.6}{1}{0,0}  \csymbol{U[\mathnormal{l}]}\thispoint{nup}{0,3}  \jointbnover{n}{0}{nup}{0}\csymbolalt[10,28]{b_\mathnormal{l}}
\end{Compose}
\]
This corresponds to somebody who is not using the app.   We would always have to use this box for such individuals since no information is available about their past.   In this case, the integer $l$ is a temporary label used in the given venue only (since we cannot track unregistered individuals from one venue to the next).

We can also have a box for somebody who is using the app but where we do not have information about their past interactions before  time $t$.  We write
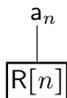
\[
\begin{Compose}{0}{0}  \setsecondfont{\mathsf}
\crectangle{n}{1.6}{1}{0,0}  \csymbol{R[\mathnormal{n}]}\thispoint{nup}{0,3}  \jointbnover{n}{0}{nup}{0}\csymbolalt[10,28]{a_\mathnormal{n}}
\end{Compose}
\]
where $t$ is the time associated with the output wire.  Such a box is useful if we can only input limited information about this individual into the calculation either for privacy reasons or because of limited computational resources.

One more box we might use is the box for a null person
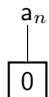
\[
\begin{Compose}{0}{0}  \setsecondfont{\mathsf}
\crectangle{n}{1}{1}{0,0}  \csymbol{0}\thispoint{nup}{0,3}  \jointbnover{n}{0}{nup}{0}\csymbolalt[10,28]{a_\mathnormal{n}}
\end{Compose}
\]
The idea here is that we may have characterised a venue with respect to more inputs into a venue than actual visitors at a given time.  Then we can feed the null input into the inputs that would otherwise be empty so that the diagram is well formed.

We also have an initialisation box for venues
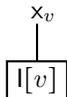
\[
\begin{Compose}{0}{0}  \setsecondfont{\mathsf}
\crectangle{n}{1.6}{1}{0,0}  \csymbol{I[\mathnormal{v}]}\thispoint{nup}{0,3}  \jointbnoarrowthick{n}{0}{nup}{0}\csymbolalt[10,28]{x_\mathnormal{v}}
\end{Compose}
\]
corresponding to when they first install the app.    We can have a similar initialisation box for an unmanaged venue
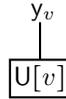
\[
\begin{Compose}{0}{0}  \setsecondfont{\mathsf}
\crectangle{n}{1.6}{1}{0,0}  \csymbol{U[\mathnormal{v}]}\thispoint{nup}{0,3}  \jointbnoarrowthick{n}{0}{nup}{0}\csymbolalt[5,28]{y_\mathnormal{v}}
\end{Compose}
\]
We feed this box into an unmanaged venue with start time $t$ associated with the outgoing wire.  The label $v$ may be drawn from map information or it could be temporary.  The box takes account of the fact that there may be harboured sources of the disease at this start time.  

\subsection{Ignore operations}

We can similarly define boxes that correspond to ignoring everything that happens to a given individual or venue after a given time.  We represent them diagrammatically as
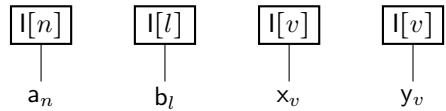
\[
\begin{Compose}{0}{0}  \setsecondfont{\mathsf}
\crectangle{n}{1.6}{1}{0,0}  \csymbol{I[\mathnormal{n}]}\thispoint{nup}{0,-3}  \csymbolalt[0,-25]{a_\mathnormal{n}}\jointbnover{nup}{0}{n}{0}
\end{Compose}
~~~~~~~
\begin{Compose}{0}{0}  \setsecondfont{\mathsf}
\crectangle{n}{1.6}{1}{0,0}  \csymbol{I[\mathnormal{l}]}\thispoint{nup}{0,-3}  \csymbolalt[0,-25]{b_\mathnormal{l}}\jointbnover{nup}{0}{n}{0}
\end{Compose}
~~~~~~~
\begin{Compose}{0}{0}  \setsecondfont{\mathsf}
\crectangle{n}{1.6}{1}{0,0}  \csymbol{I[\mathnormal{v}]}\thispoint{nup}{0,-3}  \csymbolalt[0,-25]{x_\mathnormal{v}}\jointbnover{nup}{0}{n}{0}
\end{Compose}
~~~~~~~
\begin{Compose}{0}{0}  \setsecondfont{\mathsf}
\crectangle{n}{1.6}{1}{0,0}  \csymbol{I[\mathnormal{v}]}\thispoint{nup}{0,-3}  \csymbolalt[0,-25]{y_\mathnormal{v}}\jointbnover{nup}{0}{n}{0}
\end{Compose}
\]
These boxes are useful in setting up calculations as we will see later.

\subsection{Procedures}

Venues can perform certain procedures (like cleaning surfaces) to reduce rates of infection at their venue.   We represent these by flags added to the box. For example,
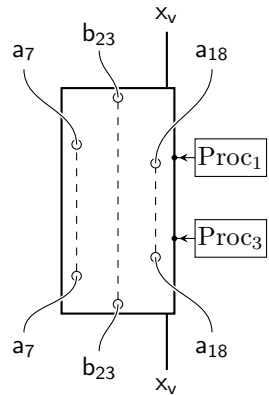
\[
\begin{Compose}{0}{0}  \setsecondfont{\mathsf} 
\crectangle{A}{3}{6}{0,0} \indashedout{A}{-2.2,-4}{-2.2,3}{7} \indashedout{A}{0,-5.5}{0,5.5}{23} \indashedout{A}{2,-3}{2,2}{18}
\relpoint{A}{2.6,-9}{vstart}\csymbolalt[0,-25]{x_v} \relpoint{A}{2.6,9}{vend} \csymbolalt[0,25]{x_v}
\relpoint{A}{-5,-7}{7start} \csymbolalt[0,-25]{a_7} \relpoint{A}{-5,7}{7end} \csymbolalt[0,25]{a_7}
\relpoint{A}{-1,-8}{23start} \csymbolalt[0,-25]{b_{23}} \relpoint{A}{-1,8}{23end} \csymbolalt[0,25]{b_{23}}
\relpoint{A}{5,-7}{18start} \csymbolalt[0,-25]{a_{18}} \relpoint{A}{5,7}{18end} \csymbolalt[0,25]{a_{18}}
\jointbnoarrowthick{vstart}{0}{A}{2.6} \jointbnoarrowthick{A}{2.6}{vend}{0}
\jointbnover{7start}{0}{A in7}{0} \jointbnover{A out7}{0}{7end}{0}
\jointbnover{23start}{0}{A in23}{0} \jointbnover{A out23}{0}{23end}{0}
\jointbnover{18start}{0}{A in18}{0} \jointbnover{A out18}{0}{18end}{0}
\rightflagdot[1.9]{A}{3,-2}{\text{Proc}_3}{3,0}
\rightflagdot[1.9]{A}{3,2.3}{\text{Proc}_1}{3,0}
\end{Compose}
\]
shows procedures $\text{Proc}_3$ and $\text{Proc}_1$ being performed.  The vertical height of the flag indicates when the procedure is performed.  The direction of the arrows indicates that these are settings (choices).

\subsection{Symptoms and test results}

Individuals can report symptoms and also test results.  We represent this by adding a flag.  The example
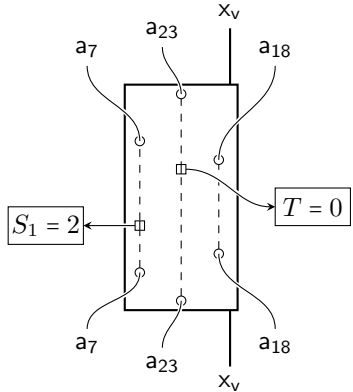
\[
\begin{Compose}{0}{0}  \setsecondfont{\mathsf} 
\crectangle{A}{3}{6}{0,0} \indashedout{A}{-2.2,-4}{-2.2,3}{7} \indashedout{A}{0,-5.5}{0,5.5}{23} \indashedout{A}{2,-3}{2,2}{18}
\relpoint{A}{2.6,-9}{vstart}\csymbolalt[0,-25]{x_v} \relpoint{A}{2.6,9}{vend} \csymbolalt[0,25]{x_v}
\relpoint{A}{-5,-7}{7start} \csymbolalt[0,-25]{a_7} \relpoint{A}{-5,7}{7end} \csymbolalt[0,25]{a_7}
\relpoint{A}{-1,-8}{23start} \csymbolalt[0,-25]{a_{23}} \relpoint{A}{-1,8}{23end} \csymbolalt[0,25]{a_{23}}
\relpoint{A}{5,-7}{18start} \csymbolalt[0,-25]{a_{18}} \relpoint{A}{5,7}{18end} \csymbolalt[0,25]{a_{18}}
\jointbnoarrowthick{vstart}{0}{A}{2.6} \jointbnoarrowthick{A}{2.6}{vend}{0}
\jointbnover{7start}{0}{A in7}{0} \jointbnover{A out7}{0}{7end}{0}
\jointbnover{23start}{0}{A in23}{0} \jointbnover{A out23}{0}{23end}{0}
\jointbnover{18start}{0}{A in18}{0} \jointbnover{A out18}{0}{18end}{0}
\leftflagsquare{A in7}{0,2.5}{S_1=2}{-5,0}
\rightflagsquare{A in23}{0,7}{T=0}{7,-2}
\end{Compose}
\]
shows individual $7$ reporting that they have symptom $S_1$ at level $2$ and individual $23$ reporting a negative test result for test $T$.  The direction of the arrows indicate that these are outcomes.

\subsection{Bluetooth and GPS}

There exist technologies to provide more tracking of peoples movement.  In the context of the COVID-19 pandemic the use of GPS and Bluetooth have been discussed \cite{googleapple2020, sattler2020risk}.   Bluetooth provides more fine-grained information than GPS and can be used to determine whether two smartphones (and thereby their owners) are proximate.  There are other technologies, such as using WiFi signals as radar to detect peoples position \cite{holl2017holography, zhao2018through} that may find application to disease control.   There are, of course, ethical considerations at play here if this is non-consensual.

We will consider the Bluetooth example here but our notation will apply to other tracking technologies.   If some of the visitors to a venue have Bluetooth tracking enabled then we will be able to collect additional information that goes beyond the behaviours those individuals have chosen.  For example, individual $7$ may choose a behaviour whereby they go to a supermarket at a certain time take a basket rather than a trolley and spend $30$ to $35$ minutes shopping.  It would be difficult to be more prescriptive than this in choosing a behaviour.   There are many trajectories around the supermarket consistent with this behaviour.   Bluetooth can provide fine-grained information on these other trajectories and, in particular, how proximate the individual is with other shoppers who have enabled Bluetooth tracking.  We take this additional information to be an outcome rather than a setting they choose (because it is unreasonable to expect individuals to decide on such prescribed trajectories).

The app can investigate relevant functions of the Bluetooth information available.     Let these functions be $R_k$ where $k=1, 2, \dots, K$ and we let them have integer values.  For example we may have the outcome $R_2=8$.  This could mean that a two given individuals were one metre apart for 10 minutes.  We add a flag to indicate these outcomes. We indicate that individuals have Bluetooth enabled while visiting a venue by using dotted rather than dashed lines.  Here is an example
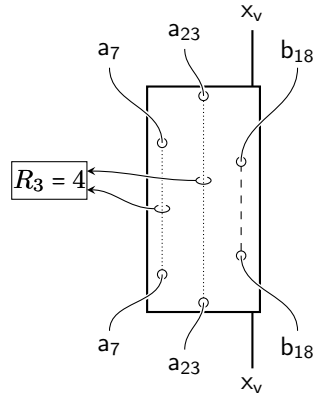
\[
\begin{Compose}{0}{0}  \setsecondfont{\mathsf} 
\crectangle{A}{3}{6}{0,0} \indottedout{A}{-2.2,-4}{-2.2,3}{7} \indottedout{A}{0,-5.5}{0,5.5}{23} \indashedout{A}{2,-3}{2,2}{18}
\relpoint{A}{2.6,-9}{vstart}\csymbolalt[0,-25]{x_v} \relpoint{A}{2.6,9}{vend} \csymbolalt[0,25]{x_v}
\relpoint{A}{-5,-7}{7start} \csymbolalt[0,-25]{a_7} \relpoint{A}{-5,7}{7end} \csymbolalt[0,25]{a_7}
\relpoint{A}{-1,-8}{23start} \csymbolalt[0,-25]{a_{23}} \relpoint{A}{-1,8}{23end} \csymbolalt[0,25]{a_{23}}
\relpoint{A}{5,-7}{18start} \csymbolalt[0,-25]{b_{18}} \relpoint{A}{5,7}{18end} \csymbolalt[0,25]{b_{18}}
\jointbnoarrowthick{vstart}{0}{A}{2.6} \jointbnoarrowthick{A}{2.6}{vend}{0}
\jointbnover{7start}{0}{A in7}{0} \jointbnover{A out7}{0}{7end}{0}
\jointbnover{23start}{0}{A in23}{0} \jointbnover{A out23}{0}{23end}{0}
\jointbnover{18start}{0}{A in18}{0} \jointbnover{A out18}{0}{18end}{0}
\flag{A in7}{-6,5}{R_3=4}\leftellipseline[->]{A in7}{3.5}{A in7}{-0.5}   \leftellipseline[->]{A in23}{6.5}{A in7}{0.5}
\end{Compose}
\]
Here we are collecting Bluetooth positioning information from individuals $7$ and $23$ and have outcome $R_3=4$.  The direction of the arrows indicates that this is an outcome.  We see that individual $18$ does not have Bluetooth enabled as they have a dashed rather than dotted line (in fact, since they are of type $\mathsf{b}$ they also are not a registered user of the app).

\subsection{Ontological interrogations}

We may wish to consider a question for which technology does not provide an actual test we can make to find the answer.   The most pertinent question is the ontological question as to whether an individual is infected with the given disease.  Any test that exists may give a false positive or negative so provides information on this  ontological question but does not fully answer it.   We represent ontological propositions by flags with dotted borders to indicate that these are not operationally available outcomes.
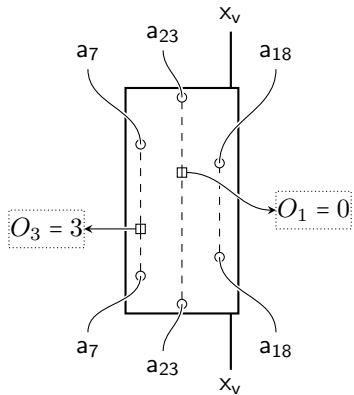
\[
\begin{Compose}{0}{0}  \setsecondfont{\mathsf} 
\crectangle{A}{3}{6}{0,0} \indashedout{A}{-2.2,-4}{-2.2,3}{7} \indashedout{A}{0,-5.5}{0,5.5}{23} \indashedout{A}{2,-3}{2,2}{18}
\relpoint{A}{2.6,-9}{vstart}\csymbolalt[0,-25]{x_v} \relpoint{A}{2.6,9}{vend} \csymbolalt[0,25]{x_v}
\relpoint{A}{-5,-7}{7start} \csymbolalt[0,-25]{a_7} \relpoint{A}{-5,7}{7end} \csymbolalt[0,25]{a_7}
\relpoint{A}{-1,-8}{23start} \csymbolalt[0,-25]{a_{23}} \relpoint{A}{-1,8}{23end} \csymbolalt[0,25]{a_{23}}
\relpoint{A}{5,-7}{18start} \csymbolalt[0,-25]{a_{18}} \relpoint{A}{5,7}{18end} \csymbolalt[0,25]{a_{18}}
\jointbnoarrowthick{vstart}{0}{A}{2.6} \jointbnoarrowthick{A}{2.6}{vend}{0}
\jointbnover{7start}{0}{A in7}{0} \jointbnover{A out7}{0}{7end}{0}
\jointbnover{23start}{0}{A in23}{0} \jointbnover{A out23}{0}{23end}{0}
\jointbnover{18start}{0}{A in18}{0} \jointbnover{A out18}{0}{18end}{0}
\leftflagsquaredotted{A in7}{0,2.5}{O_3=3}{-5,0}
\rightflagsquaredotted{A in23}{0,7}{O_1=0}{7,-2}
\end{Compose}
\]
The direction of the arrow indicates that this is similar to an outcome.  By permitting the consideration of ontological propositions we can use the probabilistic calculus to be developed in Sec. \ref{sec:probabilisticcalculations} to calculate probabilities for them.  This is important as it may determine what advice individuals are given.  For example, we may infer that somebody has a high probability of actually being infected and ask them to self-isolate.

We may be interested in ontological interrogations in between venue visits.  For these we can use the following
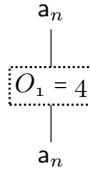
\[
\begin{Compose}{0}{0}  \setsecondfont{\mathsf}
\crectangle[thick, densely dotted]{n}{2.2}{1}{0,0}  \csymbol{\mathnormal{O_1=4}}
\thispoint{ndown}{0,-3}  \csymbolalt[0,-25]{a_\mathnormal{n}}\jointbnover{ndown}{0}{n}{0}
\thispoint{nup}{0,3}  \csymbolalt[0,25]{a_\mathnormal{n}}\jointbnover{n}{0}{nup}{0}
\end{Compose}
\]
Also consider the construction where we have an ontological interrogation followed by an ignore box
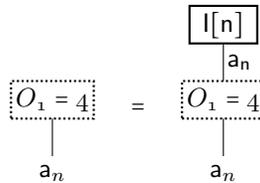
\[
\begin{Compose}{0}{0}  \setsecondfont{\mathsf}
\crectangle[thick, densely dotted]{n}{2.2}{1}{0,0}  \csymbol{\mathnormal{O_1=4}}
\thispoint{ndown}{0,-3}  \csymbolalt[0,-25]{a_\mathnormal{n}}\jointbnover{ndown}{0}{n}{0}
\end{Compose}
~~~=~~~
\begin{Compose}{0}{0}  \setsecondfont{\mathsf}
\crectangle[thick, densely dotted]{n}{2.2}{1}{0,0}  \csymbol{\mathnormal{O_1=4}}
\thispoint{ndown}{0,-3}  \csymbolalt[0,-25]{a_\mathnormal{n}}\jointbnover{ndown}{0}{n}{0}
\crectangle{nup}{1.8}{1}{0,4} \csymbol{I[\mathsf{n}]}\jointbnover[right]{n}{0}{nup}{0}\csymbolalt[10,0]{a_n}
\end{Compose}
\]
This is useful, for example,  if we are interested in whether somebody is infected after a venue visit.

We could also define ontological interrogations for venues. These might ask to what extent the venue has harboured pathogens.

\subsection{Example}\label{sec:example}

If a bunch of people visit various venues then we can represent this by wiring together operations.   For example
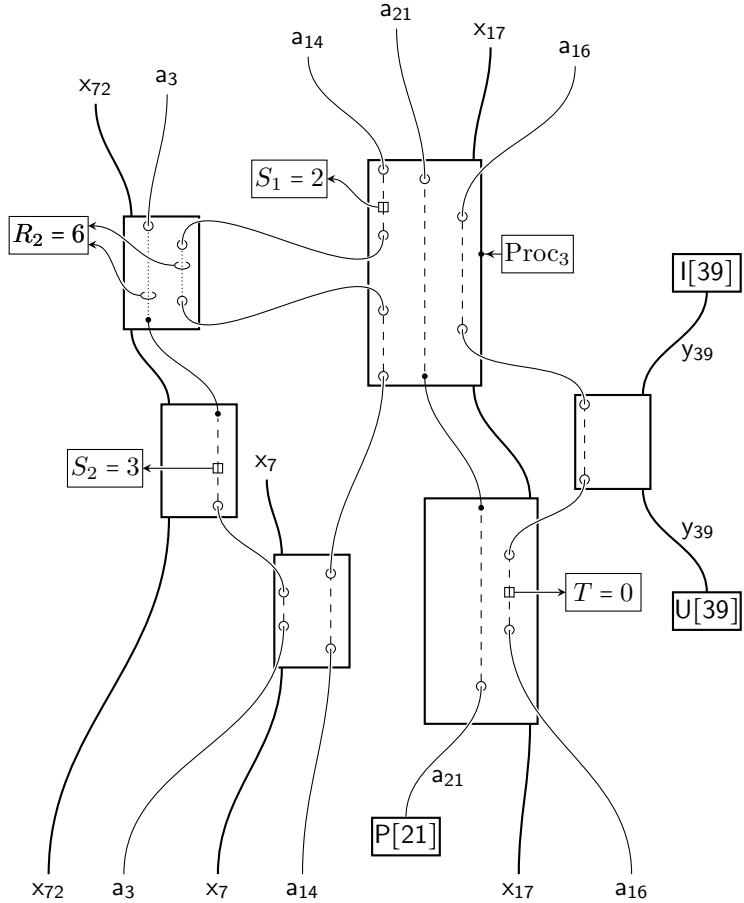
\[
\begin{Compose}{0}{0} \setsecondfont{\mathsf} 
\crectangle{A}{3}{6}{0,0} \indashedout{A}{-2.2,-5.5}{-2.2,-2}{14i}\indashedout{A}{-2.2,2}{-2.2,5.5}{14ii} \dotdashedout{A}{0,-5.5}{0,5}{21} \indashedout{A}{2,-3}{2,3}{16}
\leftflagsquare{A in14ii}{0,1.5}{S_1=2}{-5,1.5} \rightflagdot[1.9]{A}{3,1}{\text{Proc}_3}{3,0}
\crectangle{B}{2}{3}{-14,0} \dotdottedout{B}{-0.7,-2.5}{-0.7,2.5}{3} \indottedout{B}{1.1,-1.5}{1.1,1.5}{14}
\flag[2.1]{B}{-6,2}{R_2=6}\leftellipseline{B in3}{1.3}{B}{-0.5} \leftellipseline{B in14}{1.9}{B}{0.5}
\crectangle{C}{3}{6}{3,-18} \indasheddot{C}{0,-4}{0,5.5}{21} \indashedout{C}{1.5,-1}{1.5,3}{16}
\rightflagsquare{C in16}{0,2}{T=0}{5,0}
\crectangle{D}{2}{3}{-6,-18}  \indashedout{D}{-1.5,-0.8}{-1.5,1}{3} \indashedout{D}{1,-2}{1,2}{14}
\crectangle{E}{2}{3}{-12,-10} \indasheddot{E}{1,-2.4}{1,2.5}{3}\leftflagsquare[2]{E in3}{0,2}{S_2=3}{-6,0}
\thispoint{F}{-20,-32} \csymbolalt[0,-25]{x_{72}}
\thispoint{G}{-16,-32} \csymbolalt[0,-25]{a_3}
\thispoint{L}{-11,-32}  \csymbolalt[0,-25]{x_7}
\thispoint{H}{-6.5,-32} \csymbolalt[0,-25]{a_{14}}
\crectangle{I}{1.8}{1}{-1,-30} \csymbol{P[21]}
\thispoint{J}{5,-32}\csymbolalt[0,-25]{x_{17}}
\thispoint{K}{11,-32}\csymbolalt[0,-25]{a_{16}}
\crectangle{T}{2}{2.5}{10, -9}  \indashedout{T}{-1.5,-2}{-1.5,2}{16}
\crectangle{W}{1.8}{1}{15,-18}  \csymbol{U[39]}
\crectangle{I39}{1.8}{1}{15,0} \csymbol{I[39]}
\relpoint{A out14ii}{-4,6}{14end} \csymbolalt[0,25]{a_{14}}
\relpoint{A}{-1.5,13}{21end}  \csymbolalt[0,25]{a_{21}}
\relpoint{A out16}{6,8}{16end}  \csymbolalt[0,25]{a_{16}}
\relpoint{A}{3.5,12}{17end} \csymbolalt[0,25]{x_{17}}
\relpoint{B}{-3.5,9}{72end}  \csymbolalt[0,25]{x_{72}}
\relpoint{B out3}{1, 7}{3end} \csymbolalt[0,25]{a_{3}}
\relpoint{D}{-2.4, 7}{7end}  \csymbolalt[0,25]{x_{7}}   
\jointbnoarrowthick{F}{0}{E}{-1.6} \jointbnoarrowthick{E}{-1.6}{B}{-1.6} \jointbnoarrowthick{B}{-1.6}{72end}{0}
\jointbnover{G}{0}{D in3}{0} \jointbnover{D out3}{0}{E in3}{0} \jointbnoarrow{E out3}{0}{B in3}{0}
\jointbnover{B out3}{0}{3end}{0}
\jointbnoarrowthick{L}{0}{D}{-1.6} \jointbnoarrowthick{D}{-1.6}{7end}{0}
\jointbnover{H}{0}{D in14}{0}  \jointbnover{D out14}{0}{A in14i}{0} \jointbnover{A out14i}{0}{B in14}{0} \jointbnover{B out14}{0}{A in14ii}{0} \jointbnover{A out14ii}{0}{14end}{0}
\jointbnover[below right]{I}{0}{C in21}{0} \csymbolalt[-7,-25]{a_{21}} \jointbnoarrow{C out21}{0}{A in21}{0} \jointbnover{A out21}{0}{21end}{0}
\jointbnover{K}{0}{C in16}{0} \jointbnover{C out16}{0}{T in16}{0} \jointbnover{T out16}{0}{A in16}{0} \jointbnover{A out16}{0}{16end}{0}
\jointbnoarrowthick{J}{0}{C}{2.6} \jointbnoarrowthick{C}{2.6}{A}{2.6} \jointbnoarrowthick{A}{2.6}{17end}{0}
\jointbnoarrowthick[above right]{W}{0}{T}{1.6} \csymbolalt[20,0]{y_{39}}
\jointbnoarrowthick[below right]{T}{1.6}{I39}{0} \csymbolalt[20,0]{y_{39}}
\end{Compose}
\]
In this example we see how four people visit various venues for different periods of time.  Individual $3$ reports symptoms $S_2$  at level $3$ while at venue $72$.  Also, from venue $73$ Bluetooth information, $R_2=6$ is collected.   Individual $16$ receives a negative test result for test $T$ while at venue $17$.  Individual $14$ reports symptoms $S_1$ at level $2$ while in venue $17$.  Venue $17$ perform $\text{Proc}_3$ as shown.

\subsection{Circuits, fragments, preparations,  and results}\label{sec:combs}

We can construct partial pictures of peoples interactions at various venues by wiring together a bunch of boxes as in the example in Sec.\ \ref{sec:example}.   In general, there will be open wires.    It is useful to distinguish the following special cases
\begin{description}
\item[Circuits.]  If there are no open wires then we have a \emph{circuit}.
\item[Fragment.] If we have open wires then we have a \emph{fragment} (we can think of this as a fragment of a circuit).
\item[Preparation.] If we have only future pointing open wires then we have a \emph{preparation}.
\item[Results.] If we have only past pointing open wires then we have a \emph{result}.
\end{description}
Preparations and results are special cases of fragments.  A single operation is also a special case of a fragment.   We can categorise fragments further.  In particular, we can consider fragments where there are open wires in and out at intermediate times.  Here are two examples,
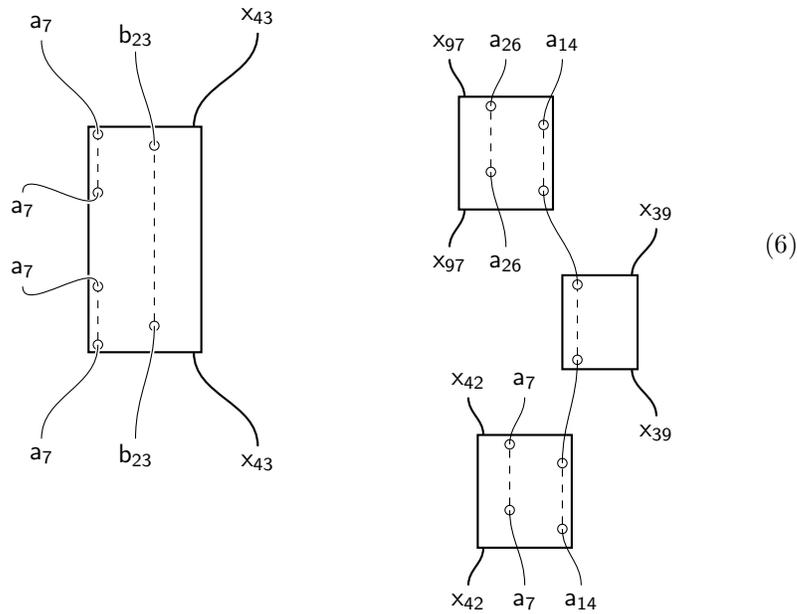
\begin{equation}\label{combexamples}
\begin{Compose}{0}{0}  \setsecondfont{\mathsf} 
\crectangle{A}{3}{6}{0,0}   \indashedout{A}{-2.5,-5.6}{-2.5,-2.5}{7i}\indashedout{A}{-2.5,2.5}{-2.5,5.6}{7ii} \indashedout{A}{0.5,-4.6}{0.5,5}{23}
\relpoint{A in7i}{-3,-5}{a7start} \csymbol[0,-25]{a_7} \relpoint{A out7i}{-4,0}{a7iend}\csymbol[0,25]{a_7}
\relpoint{A in7ii}{-4,0}{a7iistart} \csymbol[0,-25]{a_7}\relpoint{A out7ii}{-3,5}{a7end}\csymbol[0,25]{a_7}
\relpoint{A in23}{-1,-6}{b23start} \csymbol[0,-25]{b_{23}} \relpoint{A out23}{-1,5}{b23end} \csymbol[0,25]{b_{23}}
\relpoint{A}{6,-11}{x43start}\csymbol[0,-25]{x_{43}} \relpoint{A}{6,11}{x43end} \csymbol[0,25]{x_{43}}
\jointbnover{a7start}{0}{A in7i}{0} \jointbnover{A out7i}{0}{a7iend}{0} \jointbnover{a7iistart}{0}{A in7ii}{0} \jointbnover{A out7ii}{0}{a7end}{0}
\jointbnover{b23start}{0}{A in23}{0} \jointbnover{A out23}{0}{b23end}{0}
\jointbnoarrowthick{x43start}{0}{A}{2.6} \jointbnoarrowthick{A}{2.6}{x43end}{0}
\end{Compose}
~~~~~~~~~~~~~~~~
\begin{Compose}{0}{-1.1} \setdefaultfont{\mathsf} \setsecondfont{\mathsf}
\crectangle{A}{2.5}{3}{-4,-9} \indashedout{A}{-0.8,-1}{-0.8,2.5}{7} \indashedout{A}{2,-2}{2,1.5}{14}
\relpoint{A}{-3,-5}{v42start} \csymbol[0,-25]{x_{42}} \relpoint{A}{0,-5}{a7start} \csymbol[0,-25]{a_7}
\relpoint{A}{3,-5}{a14start} \csymbol[0,-25]{a_{14}}
\relpoint{A}{-3,5}{v42end} \csymbol[0,25]{x_{42}} \relpoint{A}{0,5}{a7end}\csymbol[0,25]{a_7}
\crectangle{B}{2}{2.5}{0,0} \indashedout{B}{-1.2,-2}{-1.2,2}{14}
\relpoint{B}{3,-5}{v39start} \csymbol[0,-25]{x_{39}} \relpoint{B}{3,5}{v39end}  \csymbol[0,25]{x_{39}}
\crectangle{C}{2.5}{3}{-5,9}  \indashedout{C}{-0.8,-1}{-0.8,2.5}{26} \indashedout{C}{2,-2}{2,1.5}{14}
\relpoint{C}{-3,-5}{v97start} \csymbol[0,-25]{x_{97}} \relpoint{C}{0,-5}{a26start} \csymbol[0,-25]{a_{26}}
\relpoint{C}{3,5}{a14end} \csymbol[0,25]{a_{14}}
\relpoint{C}{-3,5}{v97end} \csymbol[0,25]{x_{97}} \relpoint{C}{0,5}{a26end}\csymbol[0,25]{a_{26}}
\jointbnoarrowthick{v42start}{0}{A}{-2.2} \jointbnoarrowthick{A}{-2.2}{v42end}{0}
\jointbnoarrow{a7start}{0}{A in7}{0} \jointbnoarrow{A out7}{0}{a7end}{0}
\jointbnoarrow{a14start}{0}{A in14}{0} \jointbnoarrow{A out14}{0}{B in14}{0} \jointbnoarrow{B out14}{0}{C in14}{0} \jointbnoarrow{C out14}{0}{a14end}{0}
\jointbnoarrowthick{v39start}{0}{B}{1.7} \jointbnoarrowthick{B}{1.7}{v39end}{0}
\jointbnoarrowthick{v97start}{0}{C}{-2.2} \jointbnoarrowthick{C}{-2.2}{v97end}{0}
\jointbnoarrow{a26start}{0}{C in26}{0} \jointbnoarrow{C out26}{0}{a26end}{0}
\end{Compose}
\end{equation}
We will call these \emph{combs} in accord with similar usage in references \cite{chiribella2008quantum, chiribella2009theoretical}.  These  have gaps where pathogens (either harboured at a venue or carried by individuals can leave and come in at intermediate times. For example, somebody could leave interact with other people and then one of these other people could enter the fragment.  The example on the left shows how a single operation can have these gaps and hence count as a comb.  The combs illustrated in these examples have a single gap. We can also have combs with multiple gaps.

\subsection{Deterministic and nondeterministic fragments}\label{sec:deterministicandnondeterministicfragments}

It is worth distinguishing further.  If a fragment has outcomes on it then we say it is \emph{nondeterministic} since these outcomes do not have to happen.  On the other hand, we have no outcomes on any of the boxes comprising a fragment then it is \emph{deterministic}.
Any operation with a flag for a test, symptom, Bluetooth outcome, or an ontological interrogation is nondeterministic.  The ignore operation is deterministic.    Deterministic fragments play an important role in establishing the probabilistic theory associated with these descriptions.

There is, incidentally, an interesting subtlety concerning symptoms.  There are two ways we might monitor for symptoms: (a) we might ask people at regular intervals (once per day perhaps) if they are experiencing any symptoms, or (b) we might instruct people to report only when they do have symptoms.  In case (a) we will have outcomes like $S_2=0$ (meaning no symptoms).  In this case, an operation having no flags is deterministic.  In case (b) the lack of a symptoms flag means that the person has not experienced symptoms and, since they might have had symptoms, such we cannot then assume an operation with no flag is deterministic.   For the same of simplicity we will assume we are in case (a) in what follows though it is not difficult to deal with case (b).

\section{Probabilistic calculations using circuits}\label{sec:probabilisticcalculations}

Given the compositional description we can consider different kinds of calculation.  Here we will pursue probabilistic calculations because we want to be able to calculate costs for visiting venues and  implement probabilistic contact tracing.   We could, however, consider other kinds of calculation based on this circuit description such as deterministic contact tracing (where everybody within a certain distance of an infected individual for a certain amount of time is traced).   

\subsection{Simplified hidden variables}\label{sec:simplifiedhiddenvariables}

We denote the hidden variables associated with $\mathsf{a}_n$ by $a_n$, with $\mathsf{b}_l$ by $b_l$, with $\mathsf{x}_v$ by $x_v$ and with $\mathsf{y}_v$ by $y_v$.  For the purposes of tractability in computer modeling,  we take these to be discretized versions of these hidden variables.   As we will see shortly, these hidden variables will correspond to the indices in tensors and we will, consequently, be summing over them.

In a very simplified model we could set the hidden variables associated with a registered individual, $\mathsf{a}_n$ by
\begin{center}
\begin{tabular}{|l|l|}
\hline
$\mathsf{a}_n=0$ & \text{uninfected, not contagious, no symptoms}   \\ \hline
$\mathsf{a}_n=1$& \text{uninfected, not contagious, symptoms}   \\ \hline
$\mathsf{a}_n=2$ & \text{infected, not contagious,  no symptoms}   \\ \hline
$\mathsf{a}_n=3$ & \text{infected, not contagious, symptoms}   \\ \hline
$\mathsf{a}_n=4$ & \text{infected, contagious,  no symptoms}   \\ \hline
$\mathsf{a}_n=5$ & \text{infected,  contagious, symptoms}   \\ \hline
$\mathsf{a}_n=6$ & \text{immune, no symptoms}   \\ \hline
$\mathsf{a}_n=7$ & \text{immune, symptoms}   \\ \hline
$\mathsf{a}_n=8$ & \text{deceased}   \\ \hline
\end{tabular}
\end{center}
An unregistered individual, $\mathsf{b}_l$, could have hidden variables
\begin{center}
\begin{tabular}{|l|l|}
\hline
$\mathsf{b}_l=0$ & \text{not contagious}   \\ \hline
$\mathsf{b}_l=1$& \text{contagious}   \\ \hline
\end{tabular}
\end{center}
since, for an unregistered individuals, the app can only be concerned with the possibility they infect registered individuals.   A simplified model for the hidden variables for a venue might give only a level on some scale for the risk due to harboured sources.  For example,
\begin{center}
\begin{tabular}{|l|l|}
\hline
$\mathsf{x}_v=0$ & \text{safe}   \\ \hline
$\mathsf{x}_v=1$& \text{slight risk}   \\ \hline
$\mathsf{x}_v=2$ & \text{medium risk}   \\ \hline
$\mathsf{x}_v=3$ & \text{risky}   \\ \hline
$\mathsf{x}_v=4$ & \text{very risky}   \\ \hline
\end{tabular}
\end{center}
gives the risk on a $0$ to $4$ scale.  We can use similar hidden variables for unmanaged venues.    

This is a very simplified model.  A more sophisticated model for these hidden variables is described in Appendix \ref{app:systemsandhiddenvariables}.  

\subsection{Circuits have probabilities}

The basic assumption required to set up the mathematical framework is that we can associate a probability with a circuit where this is the joint probability of seeing all the outcomes associated with the given circuit \cite{hardy2013formalism}.  For example,
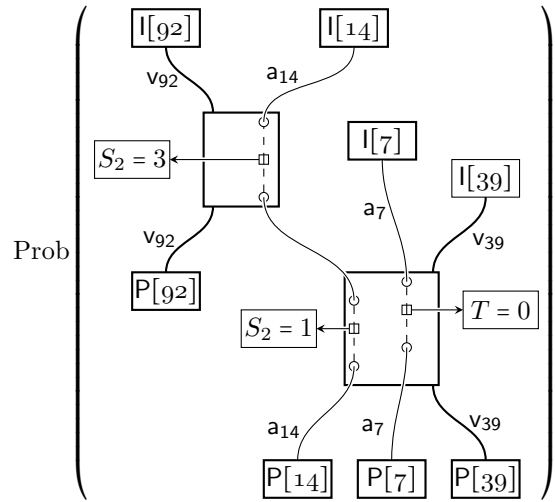
\[
\text{Prob}\left(
 \begin{Compose}{0}{-1.1} \setdefaultfont{\mathsf} \setsecondfont{\mathsf}
\crectangle{A}{2.5}{3}{0,0} \indashedout{A}{-2,-2}{-2,1.5}{14} \indashedout{A}{0.8,-1}{0.8,2.5}{7}
\crectangle{B}{2}{2.5}{-8,9} \indashedout{B}{1.2,-2}{1.2,2}{14}
\crectangle{P92}{1.8}{1}{-12,2}\csymbol{P[\mathnormal{92}]}
\crectangle{P14}{1.8}{1}{-5,-8} \csymbol{P[\mathnormal{14}]}
\crectangle{P7}{1.8}{1}{0,-8} \csymbol{P[\mathnormal{7}]}
\crectangle{P39}{1.8}{1}{5,-8} \csymbol{P[\mathnormal{39}]}
\jointbnover[left]{P14}{0}{A in14}{0} \csymbolalt[-49,-25]{a_{14}}
\jointbnover[left]{P7}{0}{A in7}{0} \csymbolalt[-25,-35]{a_{7}}
\jointbnoarrowthick[right]{P39}{0}{A}{2.2} \csymbolalt[25,0]{v_{39}}
\crectangle{E14}{1.8}{1}{-2,16} \csymbol{I[\mathnormal{14}]}
\crectangle{E92}{1.8}{1}{-12,16} \csymbol{I[\mathnormal{92}]}
\jointbnover[left]{A out14}{0}{B in14}{0} \jointbnover[above left]{B out14}{0}{E14}{0} \csymbolalt[-25,0]{a_{14}}
\crectangle{E7}{1.8}{1}{-0.5,10}\csymbol{I[\mathnormal{7}]}
\jointbnover[above left]{A out7}{0}{E7}{0}\csymbolalt[-15,0]{a_7}
\crectangle[right]{E39}{1.8}{1}{5,8} \csymbol{I[\mathnormal{39}]}
\jointbnoarrowthick[right]{A}{2.2}{E39}{0}\csymbolalt[30,-10]{v_{39}}
\jointbnoarrowthick[left]{P92}{0}{B}{-1.5} \csymbolalt[-30,0]{v_{92}}
\jointbnoarrowthick[left]{B}{-1.5}{E92}{0} \csymbolalt[-25,0]{v_{92}}
\leftflagsquare{A in14}{0,2}{S_2=1}{-4,0}
\rightflagsquare{A in7}{0,2}{T=0}{5,0}
\leftflagsquare{B in14}{0,2}{S_2=3}{-7,0}
\end{Compose}
\right)
\]
is the the joint probability that we see $S_2=1$ and $T=0$ at venue $39$ (the first operation) and $S_2=3$ at venue $92$ (the second operation).    We also assume that the probability associated with a circuit that has two disjoint parts factorises
\begin{equation}\label{disjointcircuits}
\text{Prob}(\mathsf{EF})= \text{Prob}(\mathsf{E})\text{Prob}(\mathsf{F})
\end{equation}
where circuit $\mathsf{EF}$ consists of disjoint parts $\mathsf{E}$ and $\mathsf{F}$ (each of which can be regarded as circuits in their own right).

\subsection{Ontological fiducial preparations and results}

It is useful to consider preparations that prepare in a given ontological state and results that measure onto a given ontological state.  We cannot actually implement these preparations and results in general but they play an important role in the transition from the description to the mathematics used to perform calculations.  Consider
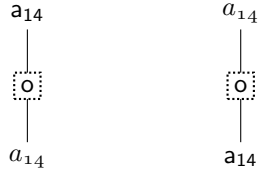
\[
\begin{Compose}{0}{-0.8}  \setdefaultfont{\mathsf} \setsecondfont{\mathnormal}
\ssquaredotted{o14}{0,0}\csymbol{o}
\relpoint{o14}{0,-3}{o14down}\csymbolalt[0,-25]{a_{14}}  \jointbnoarrow{o14down}{0}{o14}{0}
\relpoint{o14}{0,3}{o14up}\csymbol[0,25]{a_{14}} \jointbnoarrow{o14}{0}{o14up}{0}
\end{Compose}
~~~~~~~~~~~~~~~~~~
\begin{Compose}{0}{-0.8}  \setdefaultfont{\mathsf} \setsecondfont{\mathnormal}
\ssquaredotted{o14}{0,0}\csymbol{o}
\relpoint{o14}{0,-3}{o14down}\csymbol[0,-25]{a_{14}}  \jointbnoarrow{o14down}{0}{o14}{0}
\relpoint{o14}{0,3}{o14up}\csymbolalt[0,25]{a_{14}} \jointbnoarrow{o14}{0}{o14up}{0}
\end{Compose}
\]
The object on the left is the preparation of $\mathsf{a_{14}}$ in the hidden variable state $a_{14}$.   The object on the right is the result that measures to see if the hidden variable is $a_{14}$.   The \lq\lq$\mathsf{o}$" inside the box denotes \lq\lq ontological".  We call these the ontological fiducial elements in accord with terminology used elsewhere \cite{hardy2013formalism}.

\subsection{Preparations and states}

Consider the preparation $\mathsf{P}^{\mathsf{a}_{14}^1}[14]$ associated with the initialisation for individual $14$.   The superscript 1 on $a_{14}^1$ indicates that this is the first visit.   We can associate the state, $P^{a_{14}^1}$, with this as follows
\begin{equation}\label{statedefn}
\begin{Compose}{0}{-0.8}  \setsecondfont{\mathnormal}
\crectangle{n}{1.8}{1}{0,0}  \csymbol{\mathnormal{P[14]}}
\thispoint{nup}{0,4}  \csymbolalt[0,25]{a_\mathnormal{14}}\jointbnover{n}{0}{nup}{0}
\end{Compose}
~~~ =~~~
\text{Prob}\left(  ~
 \begin{Compose}{0}{-1.1} \setdefaultfont{\mathsf} \setsecondfont{\mathnormal}
\crectangle{n}{1.8}{1}{0,0}  \csymbol{P[\mathnormal{14}]}
\ssquaredotted{o14}{0,5}\csymbol{o}
\jointbnoarrow[left]{n}{0}{o14}{0} \csymbol[-15,0]{a_{14}}
\relpoint{o14}{0,3}{o14up}\csymbolalt[0,25]{a_{14}} \jointbnoarrow{o14}{0}{o14up}{0}
\end{Compose}
\right)
\end{equation}
This object, $P^{a_{14}^1}$, is the probability for the hidden variables $a_{14}$ associated with this preparation at this time.  Note that we have been using LaTeX \verb-\mathsf- font for descriptions (such as in $\mathsf{P}^{\mathsf{a}_{14}^1}[14]$).   For states (and, later, tensors) associated with these descriptions we use the \verb-\mathnormal- font of LaTeX.   As is standard in the theory of tensors, we will refer to $P^{a_{14}^1}$ as a vector even though it is really the $a_{14}^1$ component of a vector.   This vector represents the state of individual $14$ at the given time.   The lesson is that we have the following correspondence
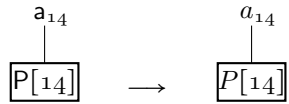
\[
 \begin{Compose}{0}{0}  \setsecondfont{\mathsf}
\crectangle{n}{1.8}{1}{0,0}  \csymbol{P[\mathnormal{14}]}\thispoint{nup}{0,3}  \jointbnover{n}{0}{nup}{0}\csymbolalt[10,28]{a_\mathnormal{14}}
\end{Compose}
~~~~\longrightarrow ~~~~
\begin{Compose}{0}{0}  \setsecondfont{\mathnormal}
\crectangle{n}{1.8}{1}{0,0}  \csymbol{\mathnormal{P[14]}}\thispoint{nup}{0,3}  \jointbnover{n}{0}{nup}{0}\csymbolalt[10,28]{a_\mathnormal{14}}
\end{Compose}
 \]
These initialisation preparations are associated with states represented by a similar picture (except for the font change).

Now consider the preparation consisting of initialisations for individuals $7$ and $14$ along with the initialisation for venue $39$.  The state is given by
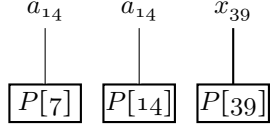
\[
\begin{Compose}{0}{0} \setdefaultfont{\mathnormal}  \setsecondfont{\mathnormal}
\crectangle{n7}{1.9}{1}{-5,0}  \csymbol{\mathnormal{P[7]}}
\thispoint{nup7}{-5,4}  \csymbolalt[0,25]{a_\mathnormal{14}}\jointbnover{n7}{0}{nup7}{0}
\crectangle{n14}{1.9}{1}{0,0}  \csymbol{\mathnormal{P[14]}}
\thispoint{nup14}{0,4}  \csymbolalt[0,25]{a_\mathnormal{14}}\jointbnover{n14}{0}{nup14}{0}
\crectangle{v39}{1.9}{1}{5,0}  \csymbol{\mathnormal{P[39]}}
\thispoint{vup39}{5,4}  \csymbolalt[0,25]{x_\mathnormal{39}}\jointbnoarrowthick{v39}{0}{vup39}{0}
\end{Compose}
\]
In symbolic notation this is
\begin{equation}
P^{a_7^1} P^{a_{14}^1} P^{x_{39}^1}
\end{equation}
The probabilities factorise at this initialisation stage.  This follows from the way we have defined states in \eqref{statedefn} and the assumption in \eqref{disjointcircuits} that probabilities over circuits with disjoint parts factorise \cite{hardy2013formalism}.   Having the probabilities factorise on initialisation might be a bad assumption if individuals $7$ and $14$ have already been spending a lot of time together at venue $39$ (this could be the house they live in or an office they share).  In this case we may wish to start with correlated initial states.   However, even if this starting assumption is wrong, its consequences would  be washed out since they would quickly become correlated according to the mathematical machinery we are now introducing.

Now consider where individuals $7$ and $14$ visit venue $39$.  This is described by the diagram
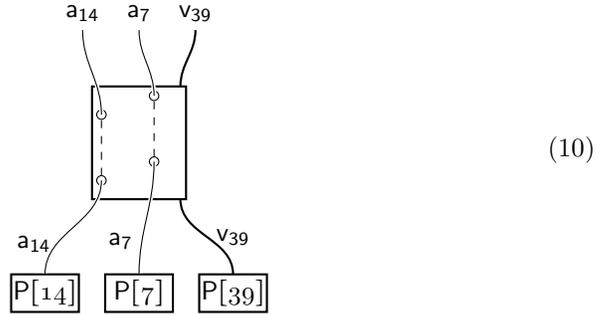
\begin{equation}\label{PPPAdiagram}
 \begin{Compose}{0}{0} \setsecondfont{\mathsf}
 \crectangle{A}{2.5}{3}{0,0} \indashedout{A}{-2,-2}{-2,1.5}{14} \indashedout{A}{0.8,-1}{0.8,2.5}{7}
 \crectangle{P14}{1.8}{1}{-5,-8} \csymbol{P[\mathnormal{14}]}
 \crectangle{P7}{1.8}{1}{0,-8} \csymbol{P[\mathnormal{7}]}
  \crectangle{P39}{1.8}{1}{5,-8} \csymbol{P[\mathnormal{39}]}
  \jointbnover[left]{P14}{0}{A in14}{0} \csymbolalt[-45,-25]{a_{14}}
  \jointbnover[left]{P7}{0}{A in7}{0} \csymbolalt[-25,-35]{a_{7}}
  \jointbnoarrowthick[right]{P39}{0}{A}{2.2} \csymbolalt[25,0]{v_{39}}
  \thispoint{E14}{-3,6} \csymbolalt[0,25]{a_{14}} \jointbnover{A out14}{0}{E14}{0}
  \thispoint{E7}{0,6} \csymbolalt[0,25]{a_7} \jointbnover{A out7}{0}{E7}{0}
  \thispoint{E39}{3,6} \csymbolalt[0,25]{v_{39}} \jointbnoarrowthick{A}{2.2}{E39}{0}
  \end{Compose}
\end{equation}
We can represent this symbolically by
\begin{equation}\label{PPPAsymbolic}
\mathsf{P}^{\mathsf{a}_7^1} \mathsf{P}^{\mathsf{a}_{14}^1} \mathsf{P}^{\mathsf{x}_{39}^1} \mathsf{A}_{\mathsf{a}_{14}^1\mathsf{a}_{7}^1\mathsf{x}_{39}^1}^{\mathsf{a}_{14}^2\mathsf{a}_{7}^2\mathsf{x}_{39}^2} 
\end{equation}
How do we turn this into a calculation?  First consider the tensor
\begin{equation}\label{tensorfromoperation}
\begin{Compose}{0}{0} \setsecondfont{\mathnormal}
 \crectangle{A}{2.5}{3}{0,0} \indashedout{A}{-2,-2}{-2,1.5}{14} \indashedout{A}{0.8,-1}{0.8,2.5}{7}
 \thispoint{op14}{-4,-7}\thispoint{op7}{0,-7} \thispoint{op39}{4,-7}
 \thispoint{or14}{-4,7}\thispoint{or7}{0,7} \thispoint{or39}{4,7}
 \relpoint{op14}{0,-3}{op14start}\csymbolalt[0,-25]{a_{14}} \relpoint{or14}{0,3}{or14out}\csymbolalt[0,25]{a_{14}}
 \relpoint{op7}{0,-3}{op7start}\csymbolalt[0,-25]{a_{7}} \relpoint{or7}{0,3}{or7out}\csymbolalt[0,25]{a_{7}}
 \relpoint{op39}{0,-3}{op39start}\csymbolalt[0,-25]{x_{39}} \relpoint{or39}{0,3}{or39out}\csymbolalt[0,25]{x_{39}}
 \jointbnoarrow{op14start}{0}{op14}{0} \jointbnoarrow[left]{op14}{0}{A in14}{0}
 \jointbnoarrow{op7start}{0}{op7}{0} \jointbnoarrow[left]{op7}{0}{A in7}{0}
 \jointbnoarrowthick{op39start}{0}{op39}{0} \jointbnoarrowthick[right]{op39}{0}{A}{2.2}
 \jointbnoarrow[left]{A out14}{0}{or14}{0} \jointbnoarrow{or14}{0}{or14out}{0}
 \jointbnoarrow[left]{A out7}{0}{or7}{0} \jointbnoarrow{or7}{0}{or7out}{0}
 \jointbnoarrowthick[right]{A}{2.2}{or39}{0} \jointbnoarrowthick{or39}{0}{or39out}{0}
\end{Compose}
=~~~~
\text{Prob}\left(
\begin{Compose}{0}{0} \setsecondfont{\mathnormal}
 \crectangle{A}{2.5}{3}{0,0} \indashedout{A}{-2,-2}{-2,1.5}{14} \indashedout{A}{0.8,-1}{0.8,2.5}{7}
 \ssquaredotted{op14}{-4,-7}\ssquaredotted{op7}{0,-7} \ssquaredotted{op39}{4,-7}
 \ssquaredotted{or14}{-4,7}\ssquaredotted{or7}{0,7} \ssquaredotted{or39}{4,7}
 \relpoint{op14}{0,-3}{op14start}\csymbolalt[0,-25]{a_{14}} \relpoint{or14}{0,3}{or14out}\csymbolalt[0,25]{a_{14}}
 \relpoint{op7}{0,-3}{op7start}\csymbolalt[0,-25]{a_{7}} \relpoint{or7}{0,3}{or7out}\csymbolalt[0,25]{a_{7}}
 \relpoint{op39}{0,-3}{op39start}\csymbolalt[0,-25]{x_{39}} \relpoint{or39}{0,3}{or39out}\csymbolalt[0,25]{x_{39}}
 \jointbnoarrow{op14start}{0}{op14}{0} \jointbnoarrow[left]{op14}{0}{A in14}{0} \csymbol[-20,0]{a_{14}}
 \jointbnoarrow{op7start}{0}{op7}{0} \jointbnoarrow[left]{op7}{0}{A in7}{0} \csymbol[-15,-20]{a_{7}}
 \jointbnoarrowthick{op39start}{0}{op39}{0} \jointbnoarrowthick[right]{op39}{0}{A}{2.2} \csymbol[20,0]{x_{39}}
 \jointbnoarrow[left]{A out14}{0}{or14}{0} \csymbol[-25,10]{a_{14}} \jointbnoarrow{or14}{0}{or14out}{0}
 \jointbnoarrow[left]{A out7}{0}{or7}{0} \csymbol[-15,0]{a_{7}} \jointbnoarrow{or7}{0}{or7out}{0}
 \jointbnoarrowthick[right]{A}{2.2}{or39}{0} \csymbol[23,-7]{x_{39}} \jointbnoarrowthick{or39}{0}{or39out}{0}
\end{Compose}
\right)
\end{equation}
This object is a tensor whose entries are probabilities.  We are using diagrammatic notation for tensors as introduced by Penrose \cite{penrose1971applications}.  We continue to use a thick wire to distinguish the venue type.    The same object can be represented in symbolic notation
\begin{equation}
A_{a_{14}^1a_{7}^1x_{39}^1}^{a_{14}^2a_{7}^2x_{39}^2}
\end{equation}
where we have now introduced integers to indicate the visit number (this starts at 1 with initialisation and increments by 1 after each operation).     This tensor is the probability that the hidden variables end up as $a_{14}^2a_{7}^2x_{39}^2$ after the operation given that they started as $a_{14}^1a_{7}^1x_{39}^1$ before the operation.  Hence, the probability for hidden variables $a_{14}^2a_{7}^2x_{39}^2$ for the situation in \eqref{PPPAdiagram} and \eqref{PPPAsymbolic} is given by
\begin{equation}\label{PPPAsymboliccalc}
P^{a_{14}^1} P^{a_7^{14}} P^{x_{39}^1}  A_{a_{14}^1a_{7}^1x_{39}^1}^{a_{14}^2a_{7}^2x_{39}^2}
\end{equation}
Here we apply Einstein's summation over the repeated indices $a_{14}^1$,  $a_{7}^1$, and $x_{39}^1$.
In Penrose's diagrammatic notation, this is given by
\begin{equation}\label{PPPAdiagramcalc}
 \begin{Compose}{0}{0} \setdefaultfont{\mathnormal} \setsecondfont{\mathnormal}
 \crectangle{A}{2.5}{3}{0,0} \indashedout{A}{-2,-2}{-2,1.5}{14} \indashedout{A}{0.8,-1}{0.8,2.5}{7}
 \crectangle{P14}{1.8}{1}{-5,-8} \csymbol{P[\mathnormal{14}]}
 \crectangle{P7}{1.8}{1}{0,-8} \csymbol{P[\mathnormal{7}]}
  \crectangle{P39}{1.8}{1}{5,-8} \csymbol{P[\mathnormal{39}]}
  \jointbnover[left]{P14}{0}{A in14}{0} \csymbolalt[-45,-25]{a_{14}}
  \jointbnover[left]{P7}{0}{A in7}{0} \csymbolalt[-25,-35]{a_{7}}
  \jointbnoarrowthick[right]{P39}{0}{A}{2.2} \csymbolalt[25,0]{v_{39}}
  \thispoint{E14}{-3,6} \csymbolalt[0,25]{a_{14}} \jointbnover{A out14}{0}{E14}{0}
  \thispoint{E7}{0,6} \csymbolalt[0,25]{a_7} \jointbnover{A out7}{0}{E7}{0}
  \thispoint{E39}{3,6} \csymbolalt[0,25]{v_{39}} \jointbnoarrowthick{A}{2.2}{E39}{0}
  \end{Compose}
\end{equation}
In the diagrammatic notation the wires that go between boxes correspond to summation over the associated indices in Penrose's diagrammatic notation.  Note that we keep the small circles and dotted lines that play no role in the Penrose notation since they provide some physical interpretation of the calculation that is being done.   In the tensor calculation, these small circles and dotted lines can be thought of as part of the label of the operation (provided by $A$ in the symbolic notation).

\subsection{The ignore operation and normalisation}\label{sec:ignoreoperations}

If the preparation is deterministic then the corresponding state will be normalised so that the probabilities to add to one.    This will be true of initialisations and we can write this normalisation condition as
\begin{equation}\label{Pnormalisation}
P^{a_{14}^1} I^{a_{14}^1} = 1
\end{equation}
where $I^{a_{14}^i}$ (for any $i$) is the vector all of whose components are $1$'s.  The vector $I^{a_{14}^i}$ is the deterministic effect and is associated with the ignore operation discussed in Sec.\ \ref{sec:ignoreoperations}.   The deterministic effect is associated with the ignore operation.   This is represented
\[
 \begin{Compose}{0}{0}  \setsecondfont{\mathsf}
\crectangle{n}{1.6}{1}{0,0}  \csymbol{I[\mathnormal{n}]}\thispoint{nup}{0,-3}  \csymbolalt[0,-25]{a_\mathnormal{n}}\jointbnover{nup}{0}{n}{0}
\end{Compose}
~~~~\rightarrow ~~~~
\begin{Compose}{0}{0} \setdefaultfont{\mathnormal} \setsecondfont{\mathnormal}
\crectangle{n}{1.6}{1}{0,0}  \csymbol{I[\mathnormal{n}]}\thispoint{nup}{0,-3}  \csymbolalt[0,-25]{a_\mathnormal{n}}\jointbnover{nup}{0}{n}{0}
\end{Compose}
 \]
 where the object on the left is simply the vector whose components are all $1$'s.   The normalisation condition in \eqref{Pnormalisation} is represented diagrammatically as follows
 \[
 \text{Prob}\left(  ~
 \begin{Compose}{0}{-0.8} \setdefaultfont{\mathsf} \setsecondfont{\mathsf}
\crectangle{n}{1.8}{1}{0,0}  \csymbol{P[\mathnormal{14}]}\crectangle{nup}{1.8}{1}{0,6}  \csymbol{I[\mathnormal{14}]} \jointbnover[left]{n}{0}{nup}{0}\csymbolalt[-25,0]{a_\mathnormal{14}}
\end{Compose}
\right) 
~~~~ =~~~~~
 \begin{Compose}{0}{-0.8} \setdefaultfont{\mathnormal} \setsecondfont{\mathnormal}
\crectangle{n}{1.8}{1}{0,0}  \csymbol{P[\mathnormal{14}]}\crectangle{nup}{1.8}{1}{0,6}  \csymbol{I[\mathnormal{14}]} \jointbnover[left]{n}{0}{nup}{0}\csymbolalt[-25,0]{a_\mathnormal{14}}
\end{Compose}
~~=1
\]
If we have more than one system, we can apply the identity to each system to get the normalisation.   As an aside, it is worth mentioning that the uniqueness of the ignore effect is equivalent to the causality condition - that future choices do not influence the present \cite{chiribella2010probabilistic}.

We are also free to consider states that are  subnormalised (for example, a state $B^{a_7^5}$ having  $B^{a_{14}^5} I^{a_{14}^5}$ less than 1). Indeed, such subnormalised states are the generic situation and occur when we have outcomes.   The state
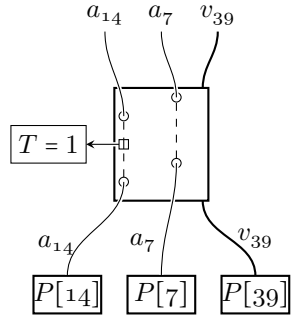
\[
\begin{Compose}{0}{0} \setdefaultfont{\mathnormal} \setsecondfont{\mathnormal}
 \crectangle{A}{2.5}{3}{0,0} \indashedout{A}{-2,-2}{-2,1.5}{14} \indashedout{A}{0.8,-1}{0.8,2.5}{7}
 \crectangle{P14}{1.8}{1}{-5,-8} \csymbol{P[\mathnormal{14}]}
 \crectangle{P7}{1.8}{1}{0,-8} \csymbol{P[\mathnormal{7}]}
  \crectangle{P39}{1.8}{1}{5,-8} \csymbol{P[\mathnormal{39}]}
  \jointbnover[left]{P14}{0}{A in14}{0} \csymbolalt[-45,-25]{a_{14}}
  \jointbnover[left]{P7}{0}{A in7}{0} \csymbolalt[-25,-35]{a_{7}}
  \jointbnoarrowthick[right]{P39}{0}{A}{2.2} \csymbolalt[25,0]{v_{39}}
  \thispoint{E14}{-3,6} \csymbolalt[0,25]{a_{14}} \jointbnover{A out14}{0}{E14}{0}
  \thispoint{E7}{0,6} \csymbolalt[0,25]{a_7} \jointbnover{A out7}{0}{E7}{0}
  \thispoint{E39}{3,6} \csymbolalt[0,25]{v_{39}} \jointbnoarrowthick{A}{2.2}{E39}{0}
   \leftflagsquare{A in14}{0,2}{T=1}{-4,0}
  \end{Compose}
\]
is not normalised since
 
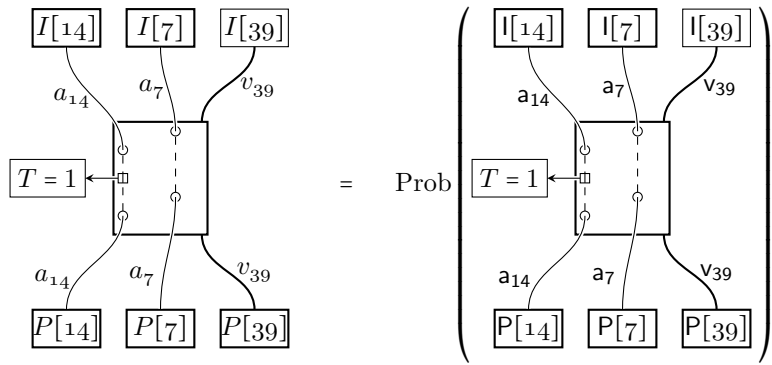
\[
 \begin{Compose}{0}{0} \setdefaultfont{\mathnormal} \setsecondfont{\mathnormal}
 \crectangle{A}{2.5}{3}{0,0} \indashedout{A}{-2,-2}{-2,1.5}{14} \indashedout{A}{0.8,-1}{0.8,2.5}{7}
 \crectangle{P14}{1.8}{1}{-5,-8} \csymbol{P[\mathnormal{14}]}
 \crectangle{P7}{1.8}{1}{0,-8} \csymbol{P[\mathnormal{7}]}
  \crectangle{P39}{1.8}{1}{5,-8} \csymbol{P[\mathnormal{39}]}
  \jointbnover[left]{P14}{0}{A in14}{0} \csymbolalt[-49,-25]{a_{14}}
  \jointbnover[left]{P7}{0}{A in7}{0} \csymbolalt[-25,-35]{a_{7}}
  \jointbnoarrowthick[right]{P39}{0}{A}{2.2} \csymbolalt[25,0]{v_{39}}
  \crectangle{E14}{1.8}{1}{-5,8} \csymbol{I[\mathnormal{14}]}
  \jointbnover[left]{A out14}{0}{E14}{0}\csymbolalt[-20,0]{a_{14}}
  \crectangle{E7}{1.8}{1}{0,8}\csymbol{I[\mathnormal{7}]} \jointbnover[left]{A out7}{0}{E7}{0}\csymbolalt[-10,0]{a_7}
  \crectangle[right]{E39}{1.8}{1}{5,8} \csymbol{I[\mathnormal{39}]}  \jointbnoarrowthick[right]{A}{2.2}{E39}{0}\csymbolalt[30,-5]{v_{39}}
    \leftflagsquare{A in14}{0,2}{T=1}{-4,0}
  \end{Compose}
  ~~~~= ~~~~
\text{Prob}\left(
 \begin{Compose}{0}{0} \setdefaultfont{\mathsf} \setsecondfont{\mathsf}
 \crectangle{A}{2.5}{3}{0,0} \indashedout{A}{-2,-2}{-2,1.5}{14} \indashedout{A}{0.8,-1}{0.8,2.5}{7}
 \crectangle{P14}{1.8}{1}{-5,-8} \csymbol{P[\mathnormal{14}]}
 \crectangle{P7}{1.8}{1}{0,-8} \csymbol{P[\mathnormal{7}]}
  \crectangle{P39}{1.8}{1}{5,-8} \csymbol{P[\mathnormal{39}]}
  \jointbnover[left]{P14}{0}{A in14}{0} \csymbolalt[-49,-25]{a_{14}}
  \jointbnover[left]{P7}{0}{A in7}{0} \csymbolalt[-25,-35]{a_{7}}
  \jointbnoarrowthick[right]{P39}{0}{A}{2.2} \csymbolalt[25,0]{v_{39}}
  \crectangle{E14}{1.8}{1}{-5,8} \csymbol{I[\mathnormal{14}]}
  \jointbnover[left]{A out14}{0}{E14}{0}\csymbolalt[-20,0]{a_{14}}
   \crectangle{E7}{1.8}{1}{0,8}\csymbol{I[\mathnormal{7}]} \jointbnover[left]{A out7}{0}{E7}{0}\csymbolalt[-10,0]{a_7}
  \crectangle[right]{E39}{1.8}{1}{5,8} \csymbol{I[\mathnormal{39}]}  \jointbnoarrowthick[right]{A}{2.2}{E39}{0}\csymbolalt[30,-4]{v_{39}}
    \leftflagsquare{A in14}{0,2}{T=1}{-4,0}
  \end{Compose}
 \right)
  \]
and this probability is equal to the probability of individual $14$ having a positive test result (and therefore less than $1$).

We can also use the ignore operation to obtain a preparation just for one system. In the example below, on the left we have a preparation for individual $7$ ignoring individual $14$ and venue $30$.   On the right we see a calculation for the state for individual $7$ which provides a probability distribution over the hidden variables $a_7$ marginalising over the hidden variables $a_{14}$ and $x_{39}$.
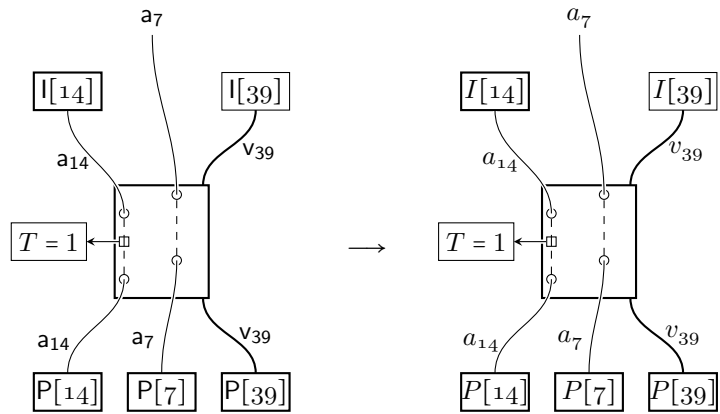
\[
 \begin{Compose}{0}{0} \setdefaultfont{\mathsf} \setsecondfont{\mathsf}
 \crectangle{A}{2.5}{3}{0,0} \indashedout{A}{-2,-2}{-2,1.5}{14} \indashedout{A}{0.8,-1}{0.8,2.5}{7}
 \crectangle{P14}{1.8}{1}{-5,-8} \csymbol{P[\mathnormal{14}]}
 \crectangle{P7}{1.8}{1}{0,-8} \csymbol{P[\mathnormal{7}]}
  \crectangle{P39}{1.8}{1}{5,-8} \csymbol{P[\mathnormal{39}]}
  \jointbnover[left]{P14}{0}{A in14}{0} \csymbolalt[-49,-25]{a_{14}}
  \jointbnover[left]{P7}{0}{A in7}{0} \csymbolalt[-25,-35]{a_{7}}
  \jointbnoarrowthick[right]{P39}{0}{A}{2.2} \csymbolalt[25,0]{v_{39}}
  \crectangle{E14}{1.8}{1}{-5,8} \csymbol{I[\mathnormal{14}]}
  \jointbnover[left]{A out14}{0}{E14}{0}\csymbolalt[-20,0]{a_{14}}
  \thispoint{E7}{-0.5,11}\csymbol[0,25]{a_7}    \jointbnover{A out7}{0}{E7}{0}
  \crectangle[right]{E39}{1.8}{1}{5,8} \csymbol{I[\mathnormal{39}]}  \jointbnoarrowthick[right]{A}{2.2}{E39}{0}\csymbolalt[30,-5]{v_{39}}
    \leftflagsquare{A in14}{0,2}{T=1}{-4,0}
  \end{Compose}
  ~~~~~\longrightarrow ~~~~~
 \begin{Compose}{0}{0} \setdefaultfont{\mathnormal} \setsecondfont{\mathnormal}
 \crectangle{A}{2.5}{3}{0,0} \indashedout{A}{-2,-2}{-2,1.5}{14} \indashedout{A}{0.8,-1}{0.8,2.5}{7}
 \crectangle{P14}{1.8}{1}{-5,-8} \csymbol{P[\mathnormal{14}]}
 \crectangle{P7}{1.8}{1}{0,-8} \csymbol{P[\mathnormal{7}]}
  \crectangle{P39}{1.8}{1}{5,-8} \csymbol{P[\mathnormal{39}]}
  \jointbnover[left]{P14}{0}{A in14}{0} \csymbolalt[-49,-25]{a_{14}}
  \jointbnover[left]{P7}{0}{A in7}{0} \csymbolalt[-25,-35]{a_{7}}
  \jointbnoarrowthick[right]{P39}{0}{A}{2.2} \csymbolalt[25,0]{v_{39}}
  \crectangle{E14}{1.8}{1}{-5,8} \csymbol{I[\mathnormal{14}]}
  \jointbnover[left]{A out14}{0}{E14}{0}\csymbolalt[-20,0]{a_{14}}
  \thispoint{E7}{-0.5,11}\csymbol[0,25]{a_7}    \jointbnover{A out7}{0}{E7}{0}
  \crectangle[right]{E39}{1.8}{1}{5,8} \csymbol{I[\mathnormal{39}]}  \jointbnoarrowthick[right]{A}{2.2}{E39}{0}\csymbolalt[30,-5]{v_{39}}
    \leftflagsquare{A in14}{0,2}{T=1}{-4,0}
  \end{Compose}
  \]
 Note that this state is subnormalised because of the outcome.

\subsection{From operations to tensors}

Generally, to convert descriptions to calculations we substitute operations with the corresponding tensors where these tensors are obtained by a calculation of the form shown in \eqref{tensorfromoperation} (that is by adding ontological fiducial elements to each wire left open and taking the probability).
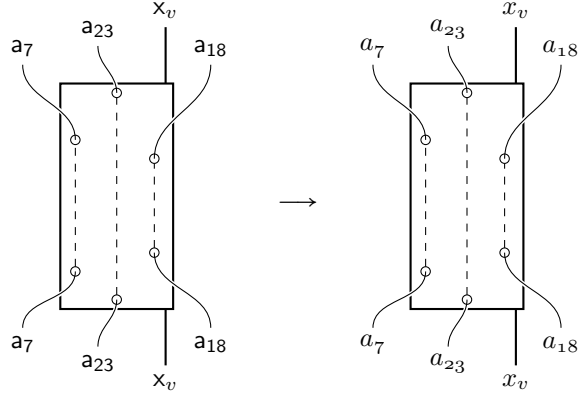
\[
\begin{Compose}{0}{0}  \setsecondfont{\mathsf} 
\crectangle{A}{3}{6}{0,0} 
\thispoint {xi}{2.6,-9} \csymbolalt[0,-25]{x_\mathnormal{v}} \jointbnoarrowthick{xi}{0}{A}{2.6}
\thispoint {xo}{2.6,9} \csymbolalt[0,25]{x_\mathnormal{v}} \joinbtnoarrowthick{xo}{0}{A}{2.6}
\thispoint{a7start}{-5,-7}\csymbolalt[0,-25]{a_7}  \thispoint{a7in}{-2.2,-4} \jointbnover{a7start}{0}{a7in}{0}
\thispoint{a7out}{-2.2,3}\thispoint{a7end}{-5,7} \csymbolalt[0,25]{a_7}
\jointbnover{a7out}{0}{a7end}{0}
\linedashedcircles{a7in}{a7out} 
\thispoint{a18start}{5,-7}\csymbolalt[0,-25]{a_{18}}  \thispoint{a18in}{2,-3} \jointbnover{a18start}{0}{a18in}{0}
\thispoint{a18out}{2,2}\thispoint{a18end}{5,7}  \csymbolalt[0,25]{a_{18}}
\jointbnover{a18out}{0}{a18end}{0}
\linedashedcircles{a18in}{a18out}  
\thispoint{a23start}{-1,-8}\csymbolalt[0,-25]{a_{23}}  \thispoint{a23in}{0,-5.5} \jointbnover{a23start}{0}{a23in}{0}
\thispoint{a23out}{0,5.5}\thispoint{a23end}{-1,8} \csymbolalt[0,25]{a_{23}}
\jointbnover{a23out}{0}{a23end}{0}
\linedashedcircles{a23in}{a23out}  
\end{Compose}
~~~~\longrightarrow~~~
\begin{Compose}{0}{0}  \setsecondfont{\mathnormal} 
\crectangle{A}{3}{6}{0,0} 
\thispoint {xi}{2.6,-9} \csymbolalt[0,-25]{x_\mathnormal{v}} \jointbnoarrowthick{xi}{0}{A}{2.6}
\thispoint {xo}{2.6,9} \csymbolalt[0,25]{x_\mathnormal{v}} \joinbtnoarrowthick{xo}{0}{A}{2.6}
\thispoint{a7start}{-5,-7}\csymbolalt[0,-25]{a_7}  \thispoint{a7in}{-2.2,-4} \jointbnover{a7start}{0}{a7in}{0}
\thispoint{a7out}{-2.2,3}\thispoint{a7end}{-5,7} \csymbolalt[0,25]{a_7}
\jointbnover{a7out}{0}{a7end}{0}
\linedashedcircles{a7in}{a7out} 
\thispoint{a18start}{5,-7}\csymbolalt[0,-25]{a_{18}}  \thispoint{a18in}{2,-3} \jointbnover{a18start}{0}{a18in}{0}
\thispoint{a18out}{2,2}\thispoint{a18end}{5,7}  \csymbolalt[0,25]{a_{18}}
\jointbnover{a18out}{0}{a18end}{0}
\linedashedcircles{a18in}{a18out}  
\thispoint{a23start}{-1,-8}\csymbolalt[0,-25]{a_{23}}  \thispoint{a23in}{0,-5.5} \jointbnover{a23start}{0}{a23in}{0}
\thispoint{a23out}{0,5.5}\thispoint{a23end}{-1,8} \csymbolalt[0,25]{a_{23}}
\jointbnover{a23out}{0}{a23end}{0}
\linedashedcircles{a23in}{a23out}  
\end{Compose}
\]
 Symbolically this is the replacement
\[
\mathsf{
{A}^{{x}_{33}^9{a}_7^{32}{a}_7^{34}{a}_{18}^{58}  }      _{{x}_{33}^8{a}_7^{31}{a}_7^{33}{a}_{18}^{57}} }
\longrightarrow
{A}^{{x}_{33}^9{a}_7^{32}{a}_7^{34}{a}_{18}^{58}  }      _{{x}_{33}^8{a}_7^{31}{a}_7^{33}{a}_{18}^{57}}
\]
The object on the right is a description. Here we use the LaTeX font \verb-\mathsf- for both the $\mathsf{A}$ symbol and in the subscripts and superscripts is a description.  The object on the left is a tensor and there \verb-\mathnormal- font is used for both the $A$ symbol and in the subscripts and superscripts. These subscripts and superscripts correspond to hidden variables associated with people and with venues.

We use the usual Einstein summation convention for repeated indices in tensor calculations.  For example, the example in \eqref{justpoppingout} corresponds to the tensor sum
\begin{equation}
{A}^{{x}_{33}^9{a}_7^{32}{a}_7^{34}{a}_{18}^{58}  }      _{{x}_{33}^8{a}_7^{31}{a}_7^{33}{a}_{18}^{57}}
{B}^{{x}_{27}^{19}{a}_7^{33}}_{{x}_{27}^{18}{a}_7^{32} }
\end{equation}
Here the repeated indices are $a_7^{32}$ and $a_7^{33}$ and these are summed over.   We can represent the summation over repeated indices in diagrammatic notation as follows (this follows from \eqref{poppingoutdiagram})
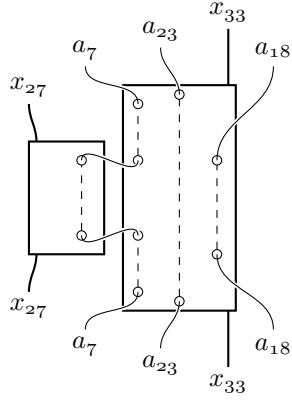
\[
\begin{Compose}{0}{0}  \setsecondfont{\mathnormal} 
\crectangle{A}{3}{6}{0,0}   \crectangle{B}{2}{3}{-6,0}
\thispoint {xi}{2.6,-9} \csymbolalt[0,-25]{x_{33}} \jointbnoarrowthick{xi}{0}{A}{2.6}
\thispoint {xo}{2.6,9} \csymbolalt[0,25]{x_{33}} \joinbtnoarrowthick{xo}{0}{A}{2.6}
\thispoint {xiB}{-8,-5} \csymbolalt[0,-25]{x_{27}} \jointbnoarrowthick{xiB}{0}{B}{-1.6}
\thispoint {xoB}{-8,5} \csymbolalt[0,25]{x_{27}} \joinbtnoarrowthick{xoB}{0}{B}{-1.6}
\thispoint{a7start}{-5,-7}\csymbolalt[0,-25]{a_7}  \thispoint{a7in}{-2.2,-5} \jointbnover{a7start}{0}{a7in}{0}
\thispoint{a7out}{-2.2,5}\thispoint{a7end}{-5,7} \csymbolalt[0,25]{a_7}
\jointbnover{a7out}{0}{a7end}{0}
\thispoint{a7leave}{-2.2, -2}  \thispoint{a7return}{-2.2,2}
\linedashedcircles{a7in}{a7leave} \linedashedcircles{a7return}{a7out}
\thispoint{a7Bin}{-5.2,-2} \thispoint{a7Bout}{-5.2,2} \jointbnover{a7leave}{0}{a7Bin}{0}  \jointbnover{a7Bout}{0}{a7return}{0} \linedashedcircles{a7Bin}{a7Bout}
\thispoint{a18start}{5,-7}\csymbolalt[0,-25]{a_{18}}  \thispoint{a18in}{2,-3} \jointbnover{a18start}{0}{a18in}{0}
\thispoint{a18out}{2,2}\thispoint{a18end}{5,7}  \csymbolalt[0,25]{a_{18}}
\jointbnover{a18out}{0}{a18end}{0}
\linedashedcircles{a18in}{a18out}  
\thispoint{a23start}{-1,-8}\csymbolalt[0,-25]{a_{23}}  \thispoint{a23in}{0,-5.5} \jointbnover{a23start}{0}{a23in}{0}
\thispoint{a23out}{0,5.5}\thispoint{a23end}{-1,8} \csymbolalt[0,25]{a_{23}}
\jointbnover{a23out}{0}{a23end}{0}
\linedashedcircles{a23in}{a23out}  
\end{Compose}
\]
once again, the font used for the indices indicates that this is a tensor calculation rather than a compositional description.  The indices associated with the wires connecting the two boxes are summed over.

\subsection{Sequential boxes and extended hidden variables}\label{sec:sequentialboxesandextendedhv}

When an individual remains in a venue as represented by sequential boxes in the diagrammatic notation then in general, as discussed in Sec. \ref{sec:sequentialboxes}, the behaviours will not factorize.  The behaviour must be specified over the two boxes.   This is denoted in diagrams by black dots.  To perform a probabilistic calculation we need to take into account that, at the intermediate time, the individual could be in many different spacial positions.   Let the position, $\mathbf{q}_n$, of individual $n$ up to some discretization be
\begin{equation}
\text{position}~~~~ \mathbf{q} \in \mathcal{S}^v_\text{pos}
\end{equation}
where $\mathcal{S}^v_\text{pos}$ is a discrete set of positions in the venue.   We define the \emph{extended hidden variables}
\begin{equation}
\tilde{a}_n  \in \mathcal{S}^n  \times \mathcal{S}^v_\text{pos}
\end{equation}
for person $n$ where $\mathcal{S}^n$ is the set of hidden variables, $a_n$, for individual $n$.     The calculation now works as follows
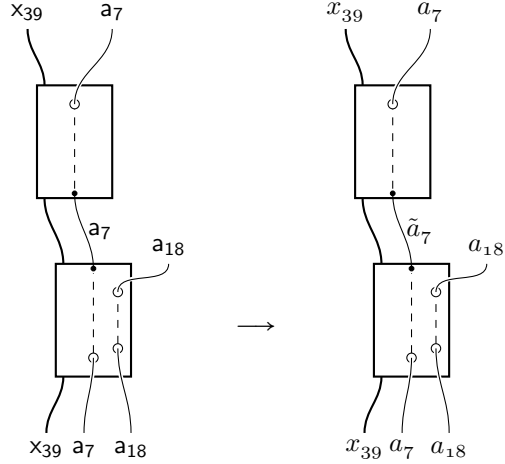
\[
\begin{Compose}{0}{0} \setsecondfont{\mathsf} 
\crectangle{A}{2}{3}{0,0} \indasheddot{A}{0,-2}{0,2.75}{7} \indashedout{A}{1.3,-1.5}{1.3,1.5}{18}
\relpoint{A}{-2.5,-6}{vstart}\csymbolalt[0,-25]{x_{39}} \relpoint{A}{-0.5,-6}{7start} \csymbolalt[0,-25]{a_7} \relpoint{A}{2,-6}{18start} \csymbolalt[0,-25]{a_{18}}
 \relpoint{A}{4,3}{18end}\csymbolalt[0,25]{a_{18}}
\crectangle{B}{2}{3}{-1,9.5} \dotdashedout{B}{0,-2.75}{0,2}{7}
\relpoint{B}{-2.5,6}{vend}\csymbolalt[0,25]{x_{39}} \relpoint{B}{2,6}{7end}\csymbolalt[0,25]{a_7}
\jointbnoarrowthick{vstart}{0}{A}{-1.6} \jointbnoarrowthick{A}{-1.6}{B}{-1.6} \jointbnoarrowthick{B}{-1.6}{vend}{0}
\jointbnover{7start}{0}{A in7}{0} \jointbnoarrow[right]{A out7}{0}{B in7}{0} \csymbolalt[10,0]{a_7} \jointbnover{B out7}{0}{7end}{0}
\jointbnover{18start}{0}{A in18}{0} \jointbnover{A out18}{0}{18end}{0}
\end{Compose}
~~~~\longrightarrow ~~~~
\begin{Compose}{0}{0} \setsecondfont{\mathnormal} 
\crectangle{A}{2}{3}{0,0} \indasheddot{A}{0,-2}{0,2.75}{7} \indashedout{A}{1.3,-1.5}{1.3,1.5}{18}
\relpoint{A}{-2.5,-6}{vstart}\csymbolalt[0,-25]{x_{39}} \relpoint{A}{-0.5,-6}{7start} \csymbolalt[0,-25]{a_7} \relpoint{A}{2,-6}{18start} \csymbolalt[0,-25]{a_{18}}
 \relpoint{A}{4,3}{18end}\csymbolalt[0,25]{a_{18}}
\crectangle{B}{2}{3}{-1,9.5} \dotdashedout{B}{0,-2.75}{0,2}{7}
\relpoint{B}{-2.5,6}{vend}\csymbolalt[0,25]{x_{39}} \relpoint{B}{2,6}{7end}\csymbolalt[0,25]{a_7}
\jointbnoarrowthick{vstart}{0}{A}{-1.6} \jointbnoarrowthick{A}{-1.6}{B}{-1.6} \jointbnoarrowthick{B}{-1.6}{vend}{0}
\jointbnover{7start}{0}{A in7}{0} \jointbnoarrow[right]{A out7}{0}{B in7}{0} \csymbolalt[12,0]{\tilde{a}_7} \jointbnover{B out7}{0}{7end}{0}
\jointbnover{18start}{0}{A in18}{0} \jointbnover{A out18}{0}{18end}{0}
\end{Compose}
\]
The $\tilde{a}_7$ at the intermediate stage indicates that we have to sum over intermediate positions, $\mathbf{q}_7$, as well as intermediate values for the hidden variables, $a_7$.   Symbolically this tensor sum is given by
\[  C_{x_{39}^{56}a_7^{23}a_{18}^{91}}^{x_{39}^{57} \tilde{a}_7^{24} a_{18}^{92}}
D_{x_{39}^{57} \tilde{a}_7^{24}}^{x_{39}^{58}a_7^{25}}   \]
where the superscripts denoting the iteration number have been added.  Here we see that $\tilde{a}_7^{24}$ is repeated denoting the sum over the extended hidden variables.

The need to resort to extended hidden variables in these circumstances is a significant complication and further study is required to see if this can be mitigated in some way (perhaps through the use of Bluetooth to locate the positions of people at the intermediate times).

\subsection{Conditions on tensors}

There are conditions on tensors associated with operations.   These are to ensure that, when we calculate probabilities associated with circuits we get a number that is (a) greater than $0$ and (b) less than $1$.   To ensure the condition (a) holds we require that the entries of the tensors are themselves in the interval $[0,1]$.  To ensure condition (b) is more involved and it is possible to give an iterative set of conditions.  The simplest of these conditions is expressed as follows.
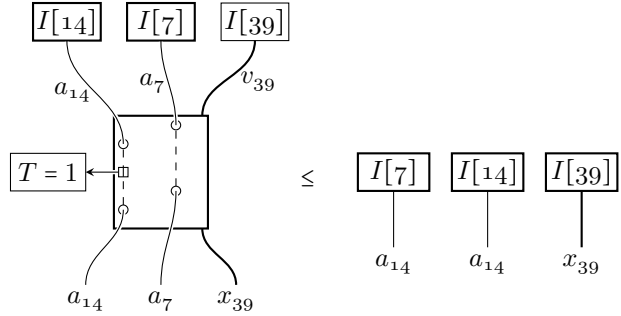
\[
\begin{Compose}{0}{0} \setdefaultfont{\mathnormal} \setsecondfont{\mathnormal}
 \crectangle{A}{2.5}{3}{0,0} \indashedout{A}{-2,-2}{-2,1.5}{14} \indashedout{A}{0.8,-1}{0.8,2.5}{7}
 \thispoint{P14}{-4,-6} \csymbol[0,-25]{a_{14}}
 \thispoint{P7}{0,-6} \csymbol[0,-25]{a_7}
 \thispoint{P39}{4,-6} \csymbol[0,-25]{x_{39}}
  \jointbnover[left]{P14}{0}{A in14}{0}
  \jointbnover[left]{P7}{0}{A in7}{0} 
  \jointbnoarrowthick[right]{P39}{0}{A}{2.2} 
  \crectangle{E14}{1.8}{1}{-5,8} \csymbol{I[\mathnormal{14}]}
  \jointbnover[left]{A out14}{0}{E14}{0}\csymbolalt[-20,0]{a_{14}}
  \crectangle{E7}{1.8}{1}{0,8}\csymbol{I[\mathnormal{7}]} \jointbnover[left]{A out7}{0}{E7}{0}\csymbolalt[-10,0]{a_7}
  \crectangle[right]{E39}{1.8}{1}{5,8} \csymbol{I[\mathnormal{39}]}  \jointbnoarrowthick[right]{A}{2.2}{E39}{0}\csymbolalt[30,-5]{v_{39}}
    \leftflagsquare{A in14}{0,2}{T=1}{-4,0}
  \end{Compose}
\leq ~~~~
\begin{Compose}{0}{0} \setdefaultfont{\mathnormal}  \setsecondfont{\mathnormal}
\crectangle{n7}{1.9}{1}{-5,0}  \csymbol{\mathnormal{I[7]}}
\thispoint{nup7}{-5,-4}  \csymbolalt[0,-25]{a_\mathnormal{14}}\jointbnover{nup7}{0}{n7}{0}
\crectangle{n14}{1.9}{1}{0,0}  \csymbol{\mathnormal{I[14]}}
\thispoint{nup14}{0,-4}  \csymbolalt[0,-25]{a_\mathnormal{14}}\jointbnover{nup14}{0}{n14}{0}
\crectangle{v39}{1.9}{1}{5,0}  \csymbol{\mathnormal{I[39]}}
\thispoint{vup39}{5,-4}  \csymbolalt[0,-25]{x_\mathnormal{39}}\jointbnoarrowthick{vup39}{0}{v39}{0}
\end{Compose}
\]
The inequality is saturated for deterministic operations.  This condition is sufficient if there are no \lq\lq gaps" as discussed for combs in Sec.\ \ref{sec:combs}.   If there are gaps  (as illustrated in the example on the left in \eqref{combexamples} where an individual leaves and then returns to the venue then certain iterative conditions also apply (the more gaps, the more conditions).  The analogous conditions have been much in the quantum case \cite{chiribella2008quantum, chiribella2009theoretical} (see also the appendix of  \cite{hardy2018construction}) and it is a simple matter to write down the corresponding conditions in the classical (rather than quantum) probability case discussed here.

\subsection{Obtaining tensors from rates}\label{sec:obtainingtensorsfromrates}

How do we calculate the tensor associated with an operation?  One way to do this is to adapt the Kolmogorov forward equation from the theory of Markov processes (see \cite{stroock2013introduction} for example).    The Kolmogorov forward equation (in notation typical for that literature) is
\begin{equation}\label{KolmogorovForwardEquation}
\frac{\partial P_{ij}(s,t)}{\partial t} = \sum_{k} P_{ik}(s,t) Q_{kj}(t)
\end{equation}
where $i$, $j$, and $k$ label the hidden variable states, $P_{ij}(s,t)$, is the probability of having jumped to state $j$ at time $t$ having been in state $i$ at time $s$, and $Q_{kj}(t)$ is the transition rate matrix.  The off-diagonal terms, $q_{kj}$, of $Q(t)$ correspond to the rate of jumping from $k$ to $j$ at time $t$.  They are taken to be positive.   For deterministic processes, if system starts at time $s$ in any given $i$ hidden variable state, then the total probability over all $j$ states it can end up in must be equal to 1 for all subsequent times $t$.  This means $\sum_j P_{ij}=1$ and consequently
\begin{equation}
\sum_j \frac{\partial P_{ij}(s,t)}{\partial t} = 0
\end{equation}
Hence,
\begin{equation}
\sum_j q_{kj} = 0 ~~~~~~~~~~~~~ q_{kk} = -\sum_{j\not= k} q_{kj}
\end{equation}
Processes that give rise to nondeterministic operations where there are outcomes  (as described in Sec.\ \ref{sec:deterministicandnondeterministicfragments}) are not treated in the literature on Markov jump processes. However, we can easily adapt the above by relaxing the conditions on the elements of $Q$.  For each outcome for a process happening over some duration (such as our operations) we have a set of possible outcomes which we will label $o=1, 2, \dots, L$.  Associated with each outcome will $P_{ij}^o(s,t)$ where this is equal to the joint probability of being in state $j$ at time $t$ and having outcome $o$ given that the state was $i$ at time $s$.  We require that
\begin{equation}
\sum_{o=1}^L P_{ij}^o(s,t)
\end{equation}
is a deterministic process (since this is the case where we ignore the outcome).   We also have transition rate matrices, $Q_{kj}^o$, associated with each outcome.   These objects satisfy the outcome-enabled Kolmogorov forward equation
\begin{equation}\label{KolmogorovForwardEquationoutcomeenabled}
\frac{\partial P_{ij}^o(s,t)}{\partial t} = \sum_{k} P_{ik}^o(s,t) Q_{kj}^o(t)
\end{equation}
The off diagonal element, $q_{kj}^o$ of $Q^o$ is the rate of transition to $jo$ for the state $k$.  Consequently the transition rate from $k$ to $j$ ignoring $o$ is
\begin{equation}
q_{kj}= \sum_o q_{kj}^o
\end{equation}
The constraints on the matrix elements of $Q^o$ are
\begin{equation}
\sum_o\sum_j q_{kj}^o = 0 ~~~~~~~~~~~~~ \sum_o q_{kk}^o = -\sum_o \sum_{j\not= k} q_{kj}^o
\end{equation}
and 
\begin{equation}
Q=\sum_o Q^o
\end{equation}
is a transition rate matrix corresponding to a deterministic process (where we ignore the outcome, $o$).  There is a subtlety in that the outcome, $o$, is something that happens over the course of the process and may not be complete at time $t$.   This means that if, up to some time $t'$, we cannot distinguish whether $o$ or $o'$ is going to happen, then we should use a single transition matrix, $Q^{o, o'}(t)$, for times $s\leq t < t'$.

We will now apply this to see how to calculate the tensors associated with operations.  First, we can rewrite any operation as a sequence  of sub-operations where each of these sub-operations has the same number of systems going in and out.  We do this by partitioning (as in Sec.\ \ref{sec:sequentialboxes}) every time an individual enters or leaves the venue.   Note this will introduce extended hidden variables such as $\tilde{a}_7$ (as discussed in Sec.\ \ref{sec:sequentialboxesandextendedhv}).  For each operation having a fixed number of systems in and out we can solve the corresponding Kolmogorov forward equation.   For example, consider calculating
\begin{equation}
A_{\tilde{a}_7^{21}\tilde{a}_{14}^{55} x_{39}^{17}}^{{a}_7^{22}\tilde{a}_{14}^{56} x_{39}^{18}}
\end{equation}
acting from time $t=T$ to $T'$.   Let the tensor associated with the operation up to time $t$ (where $T\leq t\leq T'$) be
\begin{equation}
A_{\tilde{a}_7^{21}\tilde{a}_{14}^{55} x_{39}^{17}}^{\tilde{a}_7^{22}\tilde{a}_{14}^{56} x_{39}^{18}}(T,t)
\end{equation}
We can write down the Kolmogorov forward equation as
\begin{equation}
\frac{\partial}{\partial t} A_{\tilde{a}_7^{21}\tilde{a}_{14}^{55} x_{39}^{17}}^{\tilde{a}_7^{22}\tilde{a}_{14}^{56} x_{39}^{18}}(T,t)
= A_{\tilde{a}_7^{21}\tilde{a}_{14}^{55} x_{39}^{17}}^{\tilde{a}_7^{22'}\tilde{a}_{14}^{56'} x_{39}^{18'}}(T,t)
Q_{\tilde{a}_7^{22'}\tilde{a}_{14}^{56'} x_{39}^{18'}}^{\tilde{a}_7^{22}\tilde{a}_{14}^{56} x_{39}^{18}}(t)
\end{equation}
To write down a $Q$ matrix we need a model for the spread of the disease in the given venue.   This matrix would encode
\begin{enumerate}
\item Direct transmission of the disease from one individual to another,
\item Indirect transmission of the disease whereby an individual deposits pathogens at the venue and then these infect another individual,
\item  The development of the disease in a given individual once they are infected and the decay of pathogen deposits at venues.
\end{enumerate}
Such models would  be essential for implementing the approach outlined in this paper.

\section{Using circuit techniques in counting app}

\subsection{Biggest available fragment}

When we do a calculation pertaining to an individual or venue we cannot expect to have a circuit or fragment representing the venue visits of all the people in the entire community.   Thus we need to use the biggest available fragment (BAF) to perform our calculations.  How big a fragment can be is limited by (i) privacy issues,  (ii) computational issues (iii) temporal issues.   Privacy is a factor because who visits where and  with whom is clearly sensitive information individuals may want to keep private both from government and from other people.  There is a need to consider the extent to which this information can be encrypted while still allowing the necessary calculations to be performed.  Computational issues are significant because the tensors discussed here contain a significant amount of information and this may limit the size of fragments for which calculations can be efficiently performed.  Temporal issues are important because calculations at a given time can only use information available at that time.

Given a BAF, how do we calculate a probability?   There are two approaches.  The first approach is where we close the fragment into a circuit using standard preparations, ignore results (as discussed in Sec. \ref{sec:ignoreoperations}), and ontological results.   The second approach is to use the machinery of \emph{formalism locality} whereby we look to see if the tensors associated with different fragments are proportional.  For details on this second approach see \cite{hardy2013formalism}. Here I will discuss the first approach.   Consider the fragment 
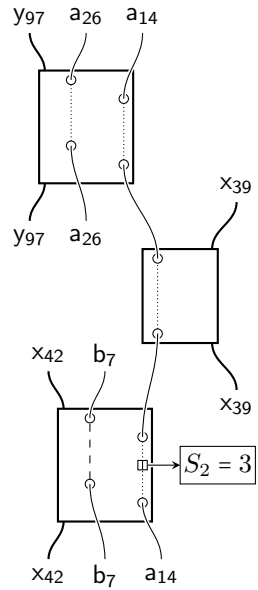
\[
\begin{Compose}{0}{-1.1} \setdefaultfont{\mathsf} \setsecondfont{\mathsf}
\crectangle{A}{2.5}{3}{-4,-9} \indashedout{A}{-0.8,-1}{-0.8,2.5}{7} \indottedout{A}{2,-2}{2,1.5}{14}
\relpoint{A}{-3,-5}{v42start} \csymbol[0,-25]{x_{42}} \relpoint{A}{0,-5}{a7start} \csymbol[0,-25]{b_7}
\relpoint{A}{3,-5}{a14start} \csymbol[0,-25]{a_{14}}
\relpoint{A}{-3,5}{v42end} \csymbol[0,25]{x_{42}} \relpoint{A}{0,5}{a7end}\csymbol[0,25]{b_7}
\crectangle{B}{2}{2.5}{0,0} \indottedout{B}{-1.2,-2}{-1.2,2}{14}
\relpoint{B}{3,-5}{v39start} \csymbol[0,-25]{x_{39}} \relpoint{B}{3,5}{v39end}  \csymbol[0,25]{x_{39}}
\crectangle{C}{2.5}{3}{-5,9}  \indottedout{C}{-0.8,-1}{-0.8,2.5}{26} \indottedout{C}{2,-2}{2,1.5}{14}
\relpoint{C}{-3,-5}{v97start} \csymbol[0,-25]{y_{97}} \relpoint{C}{0,-5}{a26start} \csymbol[0,-25]{a_{26}}
\relpoint{C}{3,5}{a14end} \csymbol[0,25]{a_{14}}
\relpoint{C}{-3,5}{v97end} \csymbol[0,25]{y_{97}} \relpoint{C}{0,5}{a26end}\csymbol[0,25]{a_{26}}
\jointbnoarrowthick{v42start}{0}{A}{-2.2} \jointbnoarrowthick{A}{-2.2}{v42end}{0}
\jointbnover{a7start}{0}{A in7}{0} \jointbnover{A out7}{0}{a7end}{0}
\jointbnover{a14start}{0}{A in14}{0} \jointbnover{A out14}{0}{B in14}{0} \jointbnover{B out14}{0}{C in14}{0} \jointbnover{C out14}{0}{a14end}{0}
\jointbnoarrowthick{v39start}{0}{B}{1.7} \jointbnoarrowthick{B}{1.7}{v39end}{0}
\jointbnoarrowthick{v97start}{0}{C}{-2.2} \jointbnoarrowthick{C}{-2.2}{v97end}{0}
\jointbnover{a26start}{0}{C in26}{0} \jointbnover{C out26}{0}{a26end}{0}
\rightflagsquare{A in14}{0,2}{S_2=3}{4,0}
\end{Compose}
\]
We would typically expect much bigger fragments in real examples. 
This fragment has a number of open wires.   We can close all these wires using the most appropriate box obtaining a circuit as follows.
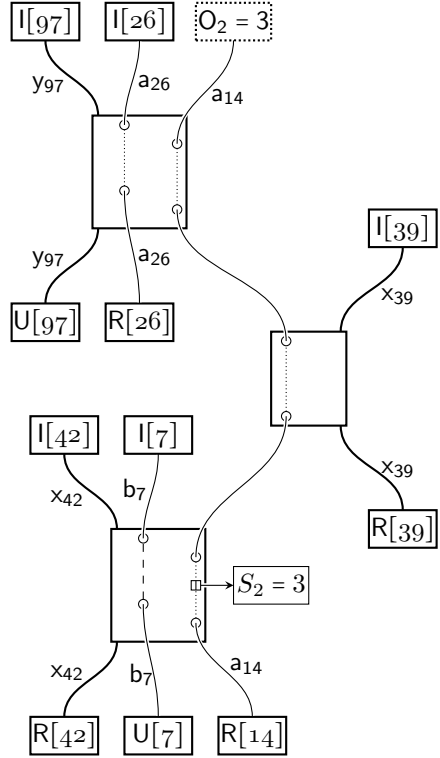
\[
 \begin{Compose}{0}{-1.1} \setdefaultfont{\mathsf} \setsecondfont{\mathsf}
\crectangle{A}{2.5}{3}{-8,-11} \indashedout{A}{-0.8,-1}{-0.8,2.5}{7} \indottedout{A}{2,-2}{2,1.5}{14}
\crectanglerelpoint{v42start}{1.8}{1}{A}{-5,-8} \csymbol{R[\mathnormal{42}]} \crectanglerelpoint{a7start}{1.8}{1}{A}{0,-8} \csymbol{U[\mathnormal{7}]}
\crectanglerelpoint{a14start}{1.8}{1}{A}{5,-8}\csymbol{R[\mathnormal{14}]}
\crectanglerelpoint{v42end}{1.8}{1}{A}{-5,8} \csymbol{I[\mathnormal{42}]} \crectanglerelpoint{a7end}{1.8}{1}{A}{0,8}\csymbol{I[\mathnormal{7}]}
\crectangle{B}{2}{2.5}{0,0} \indottedout{B}{-1.2,-2}{-1.2,2}{14}
\crectanglerelpoint{v39start}{1.8}{1}{B}{5,-8} \csymbol{R[\mathnormal{39}]} \crectanglerelpoint{v39end}{1.8}{1}{B}{5,8}  \csymbol{I[\mathnormal{39}]}
\crectangle{C}{2.5}{3}{-9,11}  \indottedout{C}{-0.8,-1}{-0.8,2.5}{26} \indottedout{C}{2,-2}{2,1.5}{14}
\crectanglerelpoint{v97start}{1.8}{1}{C}{-5,-8} \csymbol{U[\mathnormal{97}]} \crectanglerelpoint{a26start}{1.8}{1}{C}{0,-8} \csymbol{R[\mathnormal{26}]}
\crectanglerelpoint[thick, densely dotted]{a14end}{2}{1}{C}{5,8} \csymbol{O_2=3}
\crectanglerelpoint{v97end}{1.8}{1}{C}{-5,8} \csymbol{I[\mathnormal{97}]} \crectanglerelpoint{a26end}{1.8}{1}{C}{0,8}\csymbol{I[\mathnormal{26}]}
\jointbnoarrowthick[left]{v42start}{0}{A}{-2.2}\csymbol[-20,10]{x_{42}} \jointbnoarrowthick[left]{A}{-2.2}{v42end}{0}\csymbol[-20,-10]{x_{42}}
\jointbnover[left]{a7start}{0}{A in7}{0} \csymbol[0,-21]{b_7} \jointbnover[left]{A out7}{0}{a7end}{0}\csymbol[-10,10]{b_7}
\jointbnover[right]{a14start}{0}{A in14}{0}\csymbol[20,5]{a_{14}} \jointbnover{A out14}{0}{B in14}{0} \jointbnover{B out14}{0}{C in14}{0} \jointbnover[right]{C out14}{0}{a14end}{0}\csymbol[20,-7]{a_{14}}
\jointbnoarrowthick[right]{v39start}{0}{B}{1.7}\csymbol[25,-2]{x_{39}} \jointbnoarrowthick[right]{B}{1.7}{v39end}{0}\csymbol[25,-10]{x_{39}}
\jointbnoarrowthick[left]{v97start}{0}{C}{-2.2}\csymbol[-20,10]{y_{97}} \jointbnoarrowthick[left]{C}{-2.2}{v97end}{0}\csymbol[-20,-10]{y_{97}}
\jointbnover[right]{a26start}{0}{C in26}{0}\csymbol[20,-15]{a_{26}}
\jointbnover[right]{C out26}{0}{a26end}{0}\csymbol[20,0]{a_{26}}
\rightflagsquare{A in14}{0,2}{S_2=3}{4,0}
\end{Compose}
\]
Now we can calculate the probability for this circuit by replacing each operation with the corresponding tensor.

\subsection{Calculating probabilities and points}\label{sec:calculatingprobabilitiesandpoints}

In Sec.\ \ref{sec:predictedandactualcosts} we described various probabilities and points costs we are interested in calculating.  The circuit methods described in this part of the paper can be used to calculate these.     To calculate the points cost for a visit to a venue we find the biggest available fragment and calculate the increase in probability of infection between the start and the end of this visit (from which the points cost can be calculated by dividing by $p_\text{points}$).  Let $O_1=0$ if an individual is not infected and $O_1=1$ if they are infected.    For example, for individual $14$ we can calculate the increase in probability of infection on visiting venue $97$ as
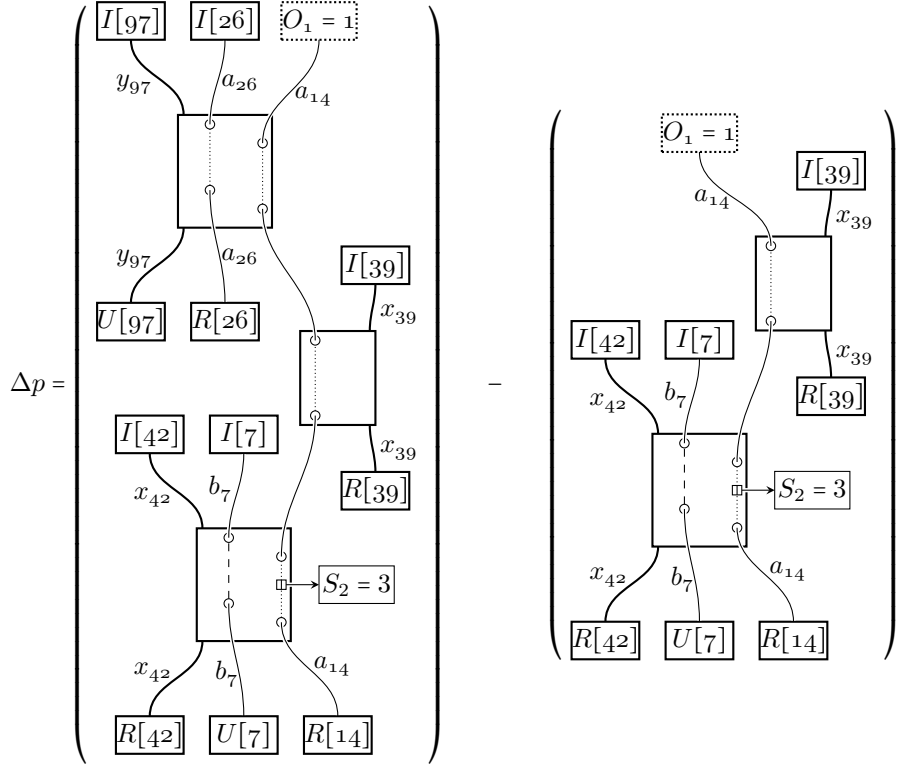
\[
\Delta p =
\left(
\begin{Compose}{0}{0} \setdefaultfont{\mathnormal} \setsecondfont{\mathnormal}
\crectangle{A}{2.5}{3}{-5,-11} \indashedout{A}{-0.8,-1}{-0.8,2.5}{7} \indottedout{A}{2,-2}{2,1.5}{14}
\crectanglerelpoint{v42start}{1.8}{1}{A}{-5,-8} \csymbol{R[\mathnormal{42}]} \crectanglerelpoint{a7start}{1.8}{1}{A}{0,-8} \csymbol{U[\mathnormal{7}]}
\crectanglerelpoint{a14start}{1.8}{1}{A}{5,-8}\csymbol{R[\mathnormal{14}]}
\crectanglerelpoint{v42end}{1.8}{1}{A}{-5,8} \csymbol{I[\mathnormal{42}]} \crectanglerelpoint{a7end}{1.8}{1}{A}{0,8}\csymbol{I[\mathnormal{7}]}
\crectangle{B}{2}{2.5}{0,0} \indottedout{B}{-1.2,-2}{-1.2,2}{14}
\crectanglerelpoint{v39start}{1.8}{1}{B}{2,-6} \csymbol{R[\mathnormal{39}]} \crectanglerelpoint{v39end}{1.8}{1}{B}{2,6}  \csymbol{I[\mathnormal{39}]}
\crectangle{C}{2.5}{3}{-6,11}  \indottedout{C}{-0.8,-1}{-0.8,2.5}{26} \indottedout{C}{2,-2}{2,1.5}{14}
\crectanglerelpoint{v97start}{1.8}{1}{C}{-5,-8} \csymbol{U[\mathnormal{97}]} \crectanglerelpoint{a26start}{1.8}{1}{C}{0,-8} \csymbol{R[\mathnormal{26}]}
\crectanglerelpoint[thick, densely dotted]{a14end}{2}{1}{C}{5,8} \csymbol{O_1=1}
\crectanglerelpoint{v97end}{1.8}{1}{C}{-5,8} \csymbol{I[\mathnormal{97}]} \crectanglerelpoint{a26end}{1.8}{1}{C}{0,8}\csymbol{I[\mathnormal{26}]}
\jointbnoarrowthick[left]{v42start}{0}{A}{-2.2}\csymbol[-20,10]{x_{42}} \jointbnoarrowthick[left]{A}{-2.2}{v42end}{0}\csymbol[-20,-10]{x_{42}}
\jointbnover[left]{a7start}{0}{A in7}{0} \csymbol[0,-21]{b_7} \jointbnover[left]{A out7}{0}{a7end}{0}\csymbol[-10,10]{b_7}
\jointbnover[right]{a14start}{0}{A in14}{0}\csymbol[20,5]{a_{14}} \jointbnover{A out14}{0}{B in14}{0} \jointbnover{B out14}{0}{C in14}{0} \jointbnover[right]{C out14}{0}{a14end}{0}\csymbol[20,-7]{a_{14}}
\jointbnoarrowthick[right]{v39start}{0}{B}{1.7}\csymbol[25,-2]{x_{39}} \jointbnoarrowthick[right]{B}{1.7}{v39end}{0}\csymbol[25,-10]{x_{39}}
\jointbnoarrowthick[left]{v97start}{0}{C}{-2.2}\csymbol[-20,10]{y_{97}} \jointbnoarrowthick[left]{C}{-2.2}{v97end}{0}\csymbol[-20,-10]{y_{97}}
\jointbnover[right]{a26start}{0}{C in26}{0}\csymbol[20,-15]{a_{26}}
\jointbnover[right]{C out26}{0}{a26end}{0}\csymbol[20,0]{a_{26}}
\rightflagsquare{A in14}{0,2}{S_2=3}{4,0}
\end{Compose}
\right)
~~~~-~~~~
\left(
\begin{Compose}{0}{1.26} \setdefaultfont{\mathnormal} \setsecondfont{\mathnormal}
\crectangle{A}{2.5}{3}{-5,-11} \indashedout{A}{-0.8,-1}{-0.8,2.5}{7} \indottedout{A}{2,-2}{2,1.5}{14}
\crectanglerelpoint{v42start}{1.8}{1}{A}{-5,-8} \csymbol{R[\mathnormal{42}]} \crectanglerelpoint{a7start}{1.8}{1}{A}{0,-8} \csymbol{U[\mathnormal{7}]}
\crectanglerelpoint{a14start}{1.8}{1}{A}{5,-8}\csymbol{R[\mathnormal{14}]}
\crectanglerelpoint{v42end}{1.8}{1}{A}{-5,8} \csymbol{I[\mathnormal{42}]} \crectanglerelpoint{a7end}{1.8}{1}{A}{0,8}\csymbol{I[\mathnormal{7}]}
\crectangle{B}{2}{2.5}{0,0} \indottedout{B}{-1.2,-2}{-1.2,2}{14}
\crectanglerelpoint{v39start}{1.8}{1}{B}{2,-6} \csymbol{R[\mathnormal{39}]}
\crectanglerelpoint{v39end}{1.8}{1}{B}{2,6}  \csymbol{I[\mathnormal{39}]}
\crectanglerelpoint[thick, densely dotted]{a14end}{2}{1}{B}{-5,8} \csymbol{O_1=1}
\jointbnoarrowthick[left]{v42start}{0}{A}{-2.2}\csymbol[-20,10]{x_{42}} \jointbnoarrowthick[left]{A}{-2.2}{v42end}{0}\csymbol[-20,-10]{x_{42}}
\jointbnover[left]{a7start}{0}{A in7}{0} \csymbol[0,-21]{b_7} \jointbnover[left]{A out7}{0}{a7end}{0}\csymbol[-10,10]{b_7}
\jointbnover[right]{a14start}{0}{A in14}{0}\csymbol[20,5]{a_{14}} \jointbnover{A out14}{0}{B in14}{0}
\jointbnover[left]{B out14}{0}{a14end}{0}\csymbol[-20,0]{a_{14}}
\jointbnoarrowthick[right]{v39start}{0}{B}{1.7}\csymbol[25,-2]{x_{39}} \jointbnoarrowthick[right]{B}{1.7}{v39end}{0}\csymbol[25,-10]{x_{39}}
\rightflagsquare{A in14}{0,2}{S_2=3}{4,0}
\end{Compose}
\right)
\]
Of course, in practise, we would be using much bigger fragments.

We calculate the probabilities discussed in Sec.\ \ref{sec:predictedandactualcosts} by substituting the appropriate operation for the venue in question.  For the example above of  a visit by individual $14$ to venue $97$ we work as follows.
\begin{description}
  \item[Predicted cost.]   For the predicted cost we substitute
  
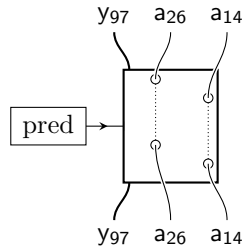
\[
 \begin{Compose}{0}{0} \setdefaultfont{\mathsf} \setsecondfont{\mathsf}
\crectangle{C}{2.5}{3}{-5,9}  \indottedout{C}{-0.8,-1}{-0.8,2.5}{26} \indottedout{C}{2,-2}{2,1.5}{14}
\relpoint{C}{-3,-5}{v97start} \csymbol[0,-25]{y_{97}}
\relpoint{C}{0,-5}{a26start} \csymbol[0,-25]{a_{26}}
\relpoint{C}{3,-5}{a14start} \csymbol[0,-25]{a_{14}}
\relpoint{C}{3,5}{a14end} \csymbol[0,25]{a_{14}}
\relpoint{C}{-3,5}{v97end} \csymbol[0,25]{y_{97}}
\relpoint{C}{0,5}{a26end}\csymbol[0,25]{a_{26}}
\jointbnoarrowthick{v97start}{0}{C}{-2.2} \jointbnoarrowthick{C}{-2.2}{v97end}{0}
\jointbnover{a26start}{0}{C in26}{0} \jointbnover{C out26}{0}{a26end}{0}
\jointbnover{a14start}{0}{C in14}{0} \jointbnover{C out14}{0}{a14end}{0}
\leftflag{C}{-2.5,0}{\text{pred}}{-4,0}
\end{Compose}
\]
into the expression for $\Delta p$.   The \lq\lq pred" flag indicates that we should use an operation describing the anticipated behaviour at the venue.   One question is what assumptions we should make in this calculation about the other visitor (labeled as $26$) to this venue.  There are various possibilities. First, they may be somebody who is known in advance (for example, somebody who works at the venue, or a previously confirmed visitor).  In this case it is reasonable to use $\mathsf{R}^{\mathsf{a}_{26}}$ as an input into this operation.  We may have even more knowledge about this person that could be expressed through a bigger BAF.  Second, the slot that $26$ occupies may not have been filled yet - so they are an unknown person.  In this case we should use an input to this node that reflects this (and takes into account the admissions policy of the venue).
  \item[Actual cost.]  When the actual visit has taken place we can use an operation that describes the actual behaviour at the venue.
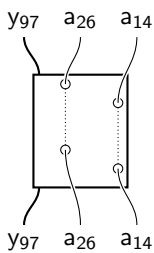
\[
  \begin{Compose}{0}{0} \setdefaultfont{\mathsf} \setsecondfont{\mathsf}
\crectangle{C}{2.5}{3}{-5,9}  \indottedout{C}{-0.8,-1}{-0.8,2.5}{26} \indottedout{C}{2,-2}{2,1.5}{14}
\relpoint{C}{-3,-5}{v97start} \csymbol[0,-25]{y_{97}}
\relpoint{C}{0,-5}{a26start} \csymbol[0,-25]{a_{26}}
\relpoint{C}{3,-5}{a14start} \csymbol[0,-25]{a_{14}}
\relpoint{C}{3,5}{a14end} \csymbol[0,25]{a_{14}}
\relpoint{C}{-3,5}{v97end} \csymbol[0,25]{y_{97}}
\relpoint{C}{0,5}{a26end}\csymbol[0,25]{a_{26}}
\jointbnoarrowthick{v97start}{0}{C}{-2.2} \jointbnoarrowthick{C}{-2.2}{v97end}{0}
\jointbnover{a26start}{0}{C in26}{0} \jointbnover{C out26}{0}{a26end}{0}
\jointbnover{a14start}{0}{C in14}{0} \jointbnover{C out14}{0}{a14end}{0}
\end{Compose}
\]
Now we can input more information about the other visitors to the venue.  Ideally, this calculation will not yield a significantly bigger probability than the predicted calculation unless the actual behaviour of individual $26$ is very different.   It is preferable for the behavioural reasons discussed in Sec.\ \ref{sec:predictedandactualcosts} not to charge individuals more points on account of matters that are beyond their control.  To deal with these issues, we could charge individual $14$ points corresponding to the minimum of the $\Delta p_\text{predicted}$ (though recalculated with individual $14$'s actual behaviour) and $\Delta p_\text{actual}$ calculated using all the actual behaviours.
\item[Fine-grained probability.]  The app is proposed to provide regular updating on the actual probability of infection based on the best information available.   This will take into account Bluetooth information, symptoms reporting, test results of the given individual and others.   To calculate this we need to use an operation with any such information available recorded on it.  For example,
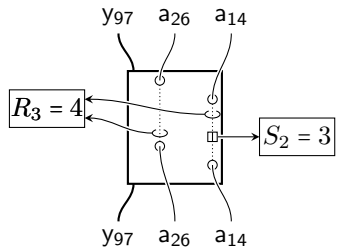
\[
\begin{Compose}{0}{0} \setdefaultfont{\mathsf} \setsecondfont{\mathsf}
\crectangle{C}{2.5}{3}{-5,9}  \indottedout{C}{-0.8,-1}{-0.8,2.5}{26} \indottedout{C}{2,-2}{2,1.5}{14}
\relpoint{C}{-3,-5}{v97start} \csymbol[0,-25]{y_{97}}
\relpoint{C}{0,-5}{a26start} \csymbol[0,-25]{a_{26}}
\relpoint{C}{3,-5}{a14start} \csymbol[0,-25]{a_{14}}
\relpoint{C}{3,5}{a14end} \csymbol[0,25]{a_{14}}
\relpoint{C}{-3,5}{v97end} \csymbol[0,25]{y_{97}}
\relpoint{C}{0,5}{a26end}\csymbol[0,25]{a_{26}}
\jointbnoarrowthick{v97start}{0}{C}{-2.2} \jointbnoarrowthick{C}{-2.2}{v97end}{0}
\jointbnover{a26start}{0}{C in26}{0} \jointbnover{C out26}{0}{a26end}{0}
\jointbnover{a14start}{0}{C in14}{0} \jointbnover{C out14}{0}{a14end}{0}
\flag{C in26}{-6,2}{R_3=4}\leftellipseline[->]{C in26}{0.7}{C in26}{-0.5}   \leftellipseline[->]{C in14}{2.7}{C in26}{0.5}
\rightflagsquare{C in14}{0,1.5}{S_2=3}{4.5,0}
\end{Compose}
\]
  \item[Updated probability.]  As new information becomes available, this can be incorporated into the calculation for the fine-grained probability possibly in the form of a bigger BAF with more information included on it.
\end{description}

\part*{Remarks and further challenges}
\addcontentsline{toc}{part}{Remarks and further challenges}

Governments have resorted to physical distancing \cite{wilder2020isolation} as the main tool to limit the spread of COVID-19.  As currently implemented it is a rather blunt tool consisting of general advice and directives to the entire population.  Physical distancing brings its own considerable risks however. First, it is not mentally healthy for people to be isolated from friends and taken out of their regular habits.  With such large numbers of people affected, this could take its own toll \cite{trout1980role}.  Second, it impedes the regular functioning of society.  This affects the provision of essential services during the crisis itself.  It also affects the longer term economic health of society with potential knock-on effects that may be enormous in the fullness of time \cite{cylus2012there, roberton2020early, magnani2020large}.  When these considerations are taken into account, the total death toll of too strong a physical distancing policy could be comparable to that due to COVID-19.   There is a need, then, to enable people to find way for people to interact to as great an extent as possible while stemming the spread of the disease.

In this paper I have described a way for people to make decisions in an epidemic to reduce their exposure and that of others.  There are many challenges on the way to implementing this idea.

Apps that solely implement contact tracing rely on the uptake to be very large before there is a significant benefit to individuals using the app.  This may not be the case for apps that enable upfront decisions such as the present proposal since then these individuals are enabled with a means to participate in actively avoiding infection for themselves and others rather than present themselves as \lq\lq sitting ducks" for infection.   There is a role for large scale simulations to prove that the points counting approach proposed here would work.  There are various questions we could ask.   Consider the case where some fraction, $\alpha$,  of people and some fraction, $\beta$, of venues participate.   (i) Is there an overall reduction in the spread of the disease for the whole population and how strongly does this depend on $\alpha$ and $\beta$? (ii) Do the individual people participating significantly reduce their own chance of infection even when the fractions $\alpha$ and $\beta$ are small? (iii) Do participating venues reduce their contribution to the spread of the disease even $\alpha$ and $\beta$ are small?   If the answer to (ii) is that the app is effective even when $\alpha$ and $\beta$ are small then there is an immediate benefit to registering and this may encourage the registration drive.

We would need to have in place a means to calculate the transition rate matrices described in Sec.\ \ref{sec:obtainingtensorsfromrates} for different venues.  This would be aided by physical models detailing how the disease spreads.  In the context of COVID 19 there has been some research concerning the physical aspects of disease spread (for example, \cite{kumar2020droplet, anfinrud2020could, evans2020avoiding, fiorillo2020covid}). It would be necessary to adapt this research to enable us to calculate the transition rate matrices.    The more sophisticated our models the more reliable they will be. However, even very primitive models (such as assuming a probability of infection function that depends on distance from an infected person and duration of exposure) would give us something to go on.   There is the possibility of using machine learning techniques to improve these models over time.  One way to use machine learning is to implement it to lean good points costings in the context of large scale simulations.

An important challenge is to ensure that the calculations that have to be done are tractable in real time.    It is worth investigating simpler models for calculating points costs (of the sort outlined in Appendix \ref{app:simplifiedpointscalc}) since these might be more tractable.  As such, a simpler approach to implement the points counting ideas in this paper would be as a supplement to an existing contact tracing app.  

Implementation of an app like this would require management and governance structures.  Setting these up goes beyond the kinds of mathematical questions discussed in this article but would be important for the practical success of such a proposal.

There are serious ethical concerns \cite{jointcivilsociety2020} with any technology that tracks peoples movements and collects sensitive information, especially if government needs to be involved.  To deal with these we need to (i) identify what the ethical issues are in any given proposal and (ii) find solutions where possible.   Task (i) requires input from experts with practical experience in ethics and policy \cite{wolff2019ethics} while task (ii) is often dealt with through cryptography.    There has been much work on cryptography for \cite{googleapple2020} for contact tracing apps and many of these ideas could be applied here also.   The challenge would be to use cryptography could be applied to hide personal identifiers, circuit information, test results, and so on, while still allowing the necessary calculations to be performed.  The probabilistic nature of the app would help provide cover for sensitive information.  For example, somebody may get a self-isolate instruction because they have interacted with one person for a long time or because they have been in the vicinity of many people for a short time.

\section*{Acknowledgements}

I am grateful to Zivy Hardy both for a comment that inspired this idea and also for proof-reading and comments on an earlier version of this article.   I am also grateful to Scott McPhee for discussions.

Research at Perimeter Institute is supported in part by the Government of Canada through the Department of Innovation, Science and Economic Development Canada and by the Province of Ontario through the Ministry of Colleges and Universities.

\part*{Appendices}

\addcontentsline{toc}{part}{Appendices}

\appendix

\section{Simplified calculation of points}\label{app:simplifiedpointscalc}

We can set up a simplified way to calculate points as illustrated here.  Let Alice have $n=1$ and Bob have $n=2$. Let $c_n^{j_n}$ be the cost to person $n$ for their $j_n$th visit to a venue since they started tracking.   In the simplified approach to calculating the cost for Alice for this cafe lunch with Bob is as follows
\begin{equation}\label{costforAlice}
  c_1^{j_1}   =   \Delta t_1\sum_{m\in V^v}  \beta^v_{u_1u_m}   S_m
\end{equation}
where the visit to the cafe is her $j_1$th visit to a venue, $\Delta t_1$ is the duration of Alice's visit, and where $V^v$ is the set of all visitors except Alice either present during Alice's visit or who have visited sufficiently recently that they may have left pathogen deposits there that are still active during Alice's visit.   The set $V^v$ includes Bob and the people working at the cafe.  Also, we define
\begin{equation}\label{Smsum}
S_m = \sum_{j_m\in J_\text{infectious}} c_m^{j_m}
\end{equation}
is the sum of the costs of previous visits to venues that individual $m$ made during such a time that they may have been infected and now be infectious (to Alice).   The label, $u_m$, indicates the behaviour of individual $m$ at the cafe (such as sitting at table number 4 from 1pm to 1:30pm).   The coefficients $\beta^v_{u_1u_m}$ are weighted according to the risks posed at venue $v$ to Alice.  These coefficients will depend on various factors such as the layout of the cafe, how well ventilated it is, and cleaning procedures.

The calculation for the cost for Alice in \eqref{costforAlice} includes a contribution, $\beta^v_{u_1u_2}   S_m$, from her interaction with Bob.  Since Bob will be sitting at the same table, the $\beta^v_{u_1u_m}$ coefficient is likely bigger than the contributions from other people at the cafe and so Alice will care more about $S_2$ for Bob.

When user $m$ first begins using the app there is some probability going in that they have the disease.   We represent this assuming that they have already accrued a certain cost, $c_m^0$.  This cost is included in the sum in \eqref{Smsum} if they have joined sufficiently recently that they may be contagious to Alice from the time before they started tracking.  The cost, $c_m^0$, can be set to
\begin{equation}\label{cm0}
c_m^0 = \frac{1}{p_\text{point}} \frac{\text{upper estimate of number of people infected when}~ m ~ \text{joins}}{\text{population}}
\end{equation}
We normalise by $p_\text{point}$ so we can work in terms of the user friendly points.   An alternative approach would be to choose $c_m^0$ to be much bigger than in \eqref{cm0}.  This would reduce the users ability to visit venues for a period of time effectively allowing a cooling off period.

A slightly more advanced approach is to substitute \eqref{Smsum} for the weighted expressions
\begin{equation}\label{Smsumgamma}
S_m = \sum_{j_m\in J}\gamma^{j_m} c_m^{j_m}
\end{equation}
where the $\gamma$'s are a measure of how contagious person $m$ during the visit to the cafe if they were infected during their earlier visit, $j_m$, to some venue.  This will depend what stage the disease is expected to have developed to in the interim.

A fraction of the population will not use the app.  In the case that such people can be identified and their behaviour modeled, they can be treated by allocating a value
\begin{equation}
S_m= k c(t)
\end{equation}
to such individuals (as used in \eqref{costforAlice}) where the constant $k$ is greater than or equal to $1$ and $c(t)$ is the fraction
\begin{equation}
c(t)   \frac{1}{p_\text{point}} \frac{\text{upper estimate of number of people infected at time}~ t}{\text{population}}
\end{equation}
This approach would treat individuals not using the app as presenting a greater risk than those that do.   Some venues may only admit people who track using the app so they can keep the infection rate for their venue below a certain recommended value.

In \eqref{Smsum} we added costs.   This is justified under certain circumstances.   Consider an individual who goes to the cafe, then to the supermarket, then to the bank.   Let the corresponding probabilities of infection for each of these stops be $p_1$, $p_2$, and $p_3$ respectively.  If each of these three chances of infection are uncorrelated then the probability of not being infected is $(1-p_1)(1-p_2)(1-p_3)$.  Hence
\begin{equation}
\text{Prob of infection} = 1 - (1-p_1)(1-p_2)(1-p_3) \approx p_1 + p_2 + p_3   ~~~\text{for}~~p_1, p_2, p_3<<1
\end{equation}
This means there is a linear regime when probabilities are small and the risks are uncorrelated.  If the probability of infection is large enough or if the situations are correlated then we cannot simply add probabilities.   In this regime, the total cost associated with these activities
\begin{equation}
\text{total cost} = c_1 + c_ 2 + c_3    ~~~ \text{where}~~~ c_i = \frac{p_i}{p_\text{point}}
\end{equation}
is approximately proportional to the probability of infection.  We will provide a simple framework for calculating the probabilities $p_i$ so that costs can be calculated.

The motivation for the approach here is that individual $m$ is infected at some earlier time then there is a certain probability he will infect Alice and so we sum up all such probabilities - this being justified when these probabilities are  small and uncorrelated.   However, both these assumptions will break down.  Probabilities will become big if somebody is showing symptoms or has a positive test result.  Probabilities will necessarily become correlated through the complex pattern of interactions of people with each other   One source of error in the formula \eqref{costforAlice} is double counting.  For example, if Alice and Bob had met several days earlier in a restaurant then there would be a contribution to Bob's total, $S_2$, from having met Alice.   This, in turn, contributes to Alice's cost when she meets Bob at the cafe.  This means we have a contribution for Alice infecting herself.  If being infected with the disease infers immunity over the timescale we are looking at then this cannot happen.  Either Alice is still infected in which case she cannot be reinfected. Or she is recovered in which case  she cannot be reinfected.   We could consider tweaking the above formulae to remove such contributions taking into account previous interactions.  One way to tweak is to go back and look at each of the $c_m^{j_m}$'s appearing in \eqref{Smsum} (or \eqref{Smsumgamma}) and see if the visit $j_m$ involved Alice.  If it did, then the calculation of $c_m^{j_m}$ can be redone setting the appropriate $\beta_{u_1u_m}^{v'}$ to zero where $v'$ is the venue for visit $j_m$.  Call this redone (or tweaked) value $c_m^{j_m}\vert_1$ (where the $1$ refers to Alice).  We can then redo the calculations of $S_m$ using $c_m^{j_m}\vert_1$ instead of $c_m^{j_m}$ and obtain the tweaked value $S_m\vert_1$.   With these tweaked values of $S_m$ we obtain a tweaked value for $c_1^{j_1}$.    These tweaked values should be used going forward.

\section{Appendix: Hidden variables}

In the main text we considered a simplified way of ascribing hidden variables to individuals and venues.  Here we consider a more sophisticated approach.   

The hidden variables, $a_n$, associated with person $n$ could detail the full biological configuration of that individual.  However, for epidemiological purposes, it is sufficient that it only details those quantities that have a bearing on the given disease.  So we can require that $a_n$ is sufficient to determine the following  appropriately discretized internal quantities
\begin{eqnarray}
  \text{infected} & & \mathbf{I}(a_n) \in \mathcal{S}_\mathbf{I} =\{0, 1, 2, \dots, L_I\} \\
  \text{contagiousness} & &  \mathbf{C}(a_n) \in  \mathcal{S}_\mathbf{C}=  \{ 0, 1, 2, \dots L_C \}   \\
  \text{susceptibility} & &  \mathbf{S}(a_n) \in \mathcal{S}_\mathbf{S}= \{ 0, 1, 2, \dots L_S \}\\
  \text{symptoms} & &   \mathbf{s}_k(a_n) \in  \mathcal{S}_{\mathbf{s}_k}= \{0, 1, 2 \dots L_{s_k} \} ~~ \text{k=1} ~\text{to}~ K  \\
  \text{alive/dead} & & \mathbf{A}(a_n) \in  \mathcal{S}_\mathbf{A}=\{1,0\}
\end{eqnarray}
where, $L_I$, $L_C$, $L_S$, and $L_{s_k}$ are integers.  To elucidate the above list a little, $\mathbf{I}(a_n)=0$ indicates that individual $n$ is uninfected and higher integers indicate a greater level of infection, $\mathbf{C}(a_n)= 4$ indicates the contagiousness level of $n$, and $\mathbf{S}_k(a_n)= 3$ indicates that individual $n$ has the symptom $k$ (for example, a cough) at level $3$.   We are taking a simplified approach here by assuming that $\mathbf{C}$ and $\mathbf{S}$ are each represented by a single number (and an integer at that) rather than being represented by a vector.  A more detailed model might require that some of these quantities are modeled by vectors. For example, the degree of contagiousness of an individual might be different through different channels (coughing, sweat deposits, \dots) and these different degrees of contagiousness could be represented by different components of a vector.   Should the vectorial aspect of these quantities be essential in the modeling, we could easily adapt approach above so that each component of the vectors is discretized (indeed, we do this below for the contagiousness of a venue).   We may want to include more quantities.  For example, the quantity $T(a_n)$, equal to the number of days since initial infection (set equal to $-1$ for the uninfected), might help modeling the development of the disease.   We might also want to include quantities relating to traces that are detected in diagnostic tests for the disease.

We can simply put
\begin{equation}
a_n \in \mathcal{S}_\mathbf{I}\times \mathcal{S}_\mathbf{C} \times \mathcal{S}_\mathbf{S} \times \mathcal{S}_{\mathbf{s}_1}\times \dots \times \mathcal{S}_{\mathbf{s}_K} \times \mathcal{S}_\mathbf{A}
\end{equation}
These quantities are related and consequently it may be possible to work with a more compressed version of the hidden variables.

For an unregistered individual our modeling need only concern itself with whether they are contagious or not and so we can put 
\begin{equation}
b_n \in  \mathcal{S}_\mathbf{C} 
\end{equation}
for such people.  

The hidden variables for a venue may give a contagiousness level for each of the harboured sources (due to pathogen deposits on different surfaces, or lingering in the air).
Let
\begin{equation}
\mathsf{C}_v(x_v)\vert_l \in \mathcal{S}_{\mathbf{C}_v\vert_l} = \{ 0, 1, \dots L_{\mathbf{C}_v\vert_l} \}
\end{equation}
where $\vert_l$ indicates the $l$th component (corresponding the $l$th harboured source).   Then we can put
\begin{equation}
x_v \in   \bigtimes_l  \mathsf{S}_{\mathbf{C}_v\vert_l}
\end{equation}
for the hidden variables (here $\bigtimes_l$ indicates the cartesian product over $l$).

\bibliography{coronabib}
\bibliographystyle{plain}

\end{document}